\documentclass[12pt,preprint]{aastex}
\newcommand{\e}{$\pm$}
\newcommand{\epeak}{{\mathcal E}_{\rm peak}}
\newcommand{\egiso}{E_{\rm \gamma,iso}}
\newcommand{\lgiso}{L_{\rm \gamma,iso}}
\newcommand{\gdur}{{\rm T}_{\rm 90,\gamma}}
\newcommand{\exiso}{E_{\rm X,iso}}
\newcommand{\lxiso}{L_{\rm X,iso}}
\newcommand{\xdur}{{\rm T}_{\rm 90,X}}
\newcommand{\eriso}{E_{\rm R,iso}}

\newcommand{\rdur}{{\rm T}_{\rm 90,R}}
\newcommand{\eoiso}{E_{\rm SN,iso}}

\newcommand{\odur}{{\rm T}_{\rm 90,SN}}
\newcommand{\ekiso}{E_{\rm k,iso}}
\newcommand{\emin}{E_{\rm min}}
\newcommand{\Msun}{M_{\odot}}
\newcommand{\rb}[1]{\raisebox{1.5ex}[0pt]{#1}}

\begin{document}

\title{Prompt and Afterglow Emission Properties of Gamma-Ray Bursts 
         with Spectroscopically Identified Supernovae}

\author{Yuki~Kaneko\altaffilmark{1},
   Enrico~Ramirez-Ruiz\altaffilmark{2,3,4},
   Jonathan~Granot\altaffilmark{5},
   Chryssa~Kouveliotou\altaffilmark{6}, 
   Stan~E.~Woosley\altaffilmark{4},
   Sandeep~K.~Patel\altaffilmark{1,6},
   Evert~Rol\altaffilmark{7},
   Jean~J.~M.~in't~Zand\altaffilmark{8},
   Alexander~J.~van~der~Horst\altaffilmark{9},
   Ralph~A.~M.~J.~Wijers\altaffilmark{9}, and
   Richard~Strom\altaffilmark{10,9}
}

\altaffiltext{1}{Universities Space Research Association,
   NSSTC, VP62, 320 Sparkman Drive, Huntsville, AL 35805,
   {\tt Yuki.Kaneko@nsstc.nasa.gov}}
\altaffiltext{2}{Institute for Advanced Study, 
   Einstein Drive, Princeton, NJ 08540}
\altaffiltext{3}{Chandra Fellow}
\altaffiltext{4}{Department of Astronomy and Astrophysics, University
of California, Santa Cruz, CA 95064}
\altaffiltext{5}{Kavli Institute for Particle Astrophysics and
   Cosmology,
Stanford University, P.O. Box 20450, MS 29, Stanford, CA 94309}
\altaffiltext{6}{NASA/MSFC, NSSTC, VP62, 320 Sparkman Drive, 
   Huntsville, AL 35805}
\altaffiltext{7}{Department of Physics and Astronomy, University of 
   Leicester, University Road, Leicester LE1 RH7, United Kingdom}
\altaffiltext{8}{SRON Netherlands Institute for Space Research, 
   Sorbonnelaan 2, 3584 CA, Utrecht, The Netherlands \&
   Astronomical Institute, Utrecht University, P.O. Box 80 000, 3508 TA, 
   Utrecht, The Netherlands}
\altaffiltext{9}{Astronomical Institute, University of Amsterdam, Kruislaan 
403, 1098SJ Amsterdam, The Netherlands}
\altaffiltext{10}{ASTRON, PO Box 2, 7990 AA Dwingeloo, The Netherlands}

\begin{abstract}

We present a detailed spectral analysis of the prompt and afterglow
emission of four nearby long-soft gamma-ray bursts 
(GRBs~980425, 030329, 031203, and 060218) that were
spectroscopically found to be associated with type Ic supernovae, and
compare them to the general GRB population. For each event, we
investigate the spectral and luminosity evolution, and estimate the
total energy budget based upon broadband observations. The
observational inventory for these events has become rich enough to
allow estimates of their energy content in relativistic and
sub-relativistic form.  The result is a global portrait of the effects
of the physical processes responsible for producing long-soft GRBs. 
In particular, we find that the values of the energy released in mildly 
relativistic outflows appears to have a significantly smaller scatter than 
those found in highly relativistic ejecta.
This is consistent with a picture in which the energy released
inside the progenitor star is roughly standard, while the fraction of
that energy that ends up in highly relativistic ejecta outside the
star can vary dramatically between different events.

\end{abstract}

\section{Introduction}

The discovery in 1998 of a Gamma-Ray Burst (GRB) in coincidence with a
very unusual supernova (SN) of Type Ic (GRB\,980425 -- SN\,1998bw; 
spectroscopically identified) was a turning point in the study of GRBs, offering 
compelling evidence that long-soft GRBs \citep{kou93} are indeed associated with the deaths 
of massive stars \citep{gal98}. The large energy release inferred for the 
supernova suggested that GRBs are potentially associated with a novel class of
explosions, having unusual properties in terms of their energy, asymmetry, 
and relativistic ejecta.  More importantly, however, GRB\,980425 provided 
the first hint that GRBs might have an intrinsic broad range of energies: 
the total energy output in $\gamma$-rays (assuming an isotropic energy 
release) was only $\egiso \approx 7 \times 10^{47}$\,erg, some four orders 
of magnitude less energy than that associated with typical GRBs \citep{blo03}.
Finally, the fact that GRB\,980425/SN\,1998bw was located in a nearby galaxy
with a redshift $z$ = 0.0085 (\citealt{tin98}; at\footnote{Throughout the paper 
we assume a cosmology with $H_0 = 72\;{\rm km\; s^{-1}\;Mpc^{-1}}$, $\Omega_{M} 
= 0.27$, and $\Omega_{\Lambda} = 0.73$.} $35.6\;$Mpc it remains the closest GRB 
to the Earth] gave rise to the possibility that such lower-energy bursts 
might be more common than had previously been thought, but harder to detect 
due to instrumental sensitivity.
 
Unfortunately, during the elapsed eight years, very few SNe have been
observed simultaneously with GRBs. To date, three more nearby GRBs have been 
unambiguously, spectroscopically identified with supernovae, two of which were 
discovered in 2003 (GRB\,030329 -- SN\,2003dh; \citealt{sta03,hjo03}, 
GRB\,031203 -- SN\,2003lw; \citealt{tag03}) and one in 2006 (GRB\,060218 -- 
SN\,2006aj; \citealt{pia06}).  Each of these SNe is of the same unusual type 
as SN\,1998bw.  Note, however, that there is weaker photometric evidence that 
many other GRBs are accompanied by SNe, mainly by identification of a late time 
``SN bump'' in the GRB optical afterglow lightcurve 
\citep[e.g.][]{blo99,gal00,zeh04,woo06}.  
In this paper, we address only the four events with spectroscopically verified 
SN associations.  Among these, only GRB\,060218 (interestingly at 143.2\,Mpc for
$z = 0.0335$, the closest after GRB\,980425; \citealt{mir06,pia06}) had 
$\gamma$-ray energetics somewhat comparable to 
GRB\,980425, and could possibly be added to the intrinsically-faint GRB sample.  
In the case of GRB\,030329, the total energy release was at the low end of
the typical range ($\egiso \sim 10^{52}\;$erg), much higher than in the other 
three events.  In fact, SN\,2003dh was obscured by the extreme optical 
brightness of the GRB afterglow and was only detected spectroscopically in the 
GRB optical lightcurve. Finally, the total energy release of GRB\,031203 was 
intermediate between that of GRB\,980425 and regular GRBs.  \citet{ram05} argued 
that the faint GRB\,031203 was a typical powerful GRB viewed at an angle 
slightly greater than about twice the half-opening angle of the central jet.
 
In the present study, we consider the energy released during the GRBs, the 
afterglow, and the SN explosion for these four events in all wavebands, from 
$\gamma$-rays to radio waves, and we also estimate their kinetic energy content.
The properties of the prompt and afterglow emission are described in detail in 
\S\S~\ref{sec:prompt} and \ref{sec:AG}, respectively, and are compared in 
\S~\ref{sec:comparison}. The bolometric energy calculation and the evolution of 
the explosion responsible for their associated SNe are presented in 
\S~\ref{sec:SN}, while \S~\ref{sec:dis} discusses the combined GRB-SN 
properties and their potential implications. We conclude in \S~\ref{sec:summary} 
with a brief summary of our primary results and their implications.

\section{Prompt Emission} \label{sec:prompt}

\subsection{GRB\,980425}

\begin{figure}
\centerline{
\plotone{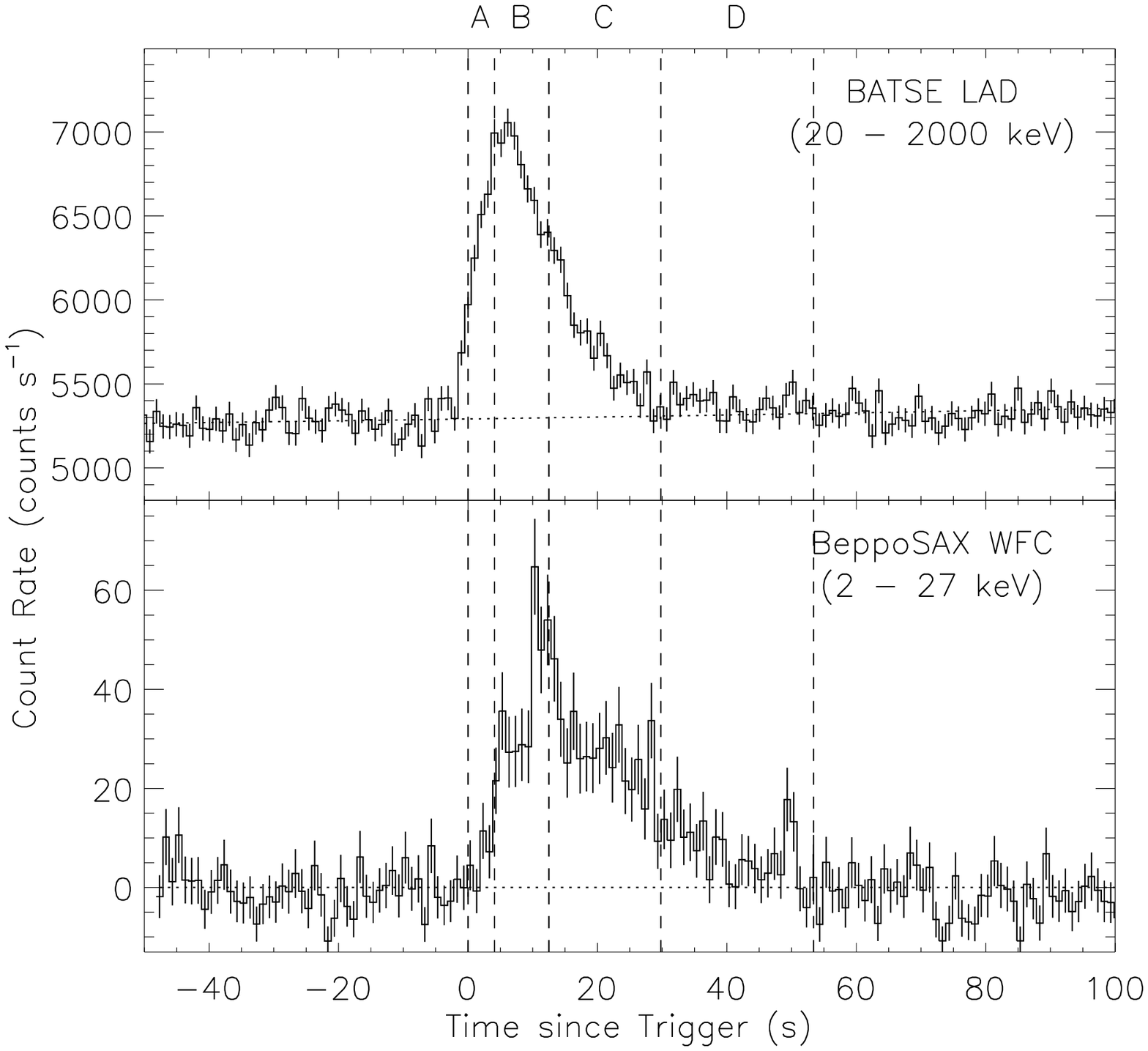}}
\caption{Lightcurves of GRB\,980425 detected with the brightest BATSE LAD (20$-$2000\,keV; 
{\it top}) and with the {\it BeppoSAX} WFC (2$-$27\,keV; {\it bottom}) plotted with 1-s 
resolution. The four time intervals used in the current analysis are labeled 
with A, B, C, and D.  The dotted lines indicate background levels.}
\label{fig:980425_lc}
\end{figure}

GRB\,980425 triggered the Large Area Detectors (LADs; 20$-$2000\,keV) of the 
Burst and Transient Source Experiment (BATSE) on board the {\it Compton 
Gamma-Ray Observatory}, on 1998 April 25 at 21:49:09 UT \citep{kip98}.  
The burst was also observed with the Gamma-Ray Burst Monitor (GRBM; 
40$-$700\,keV) and one of the two Wide Field Cameras (WFC; 2$-$27\,keV) on 
board the {\it BeppoSAX} satellite.  Figure~\ref{fig:980425_lc} shows the 
brightest BATSE/LAD and WFC lightcurves of the burst. \citet{fro00} performed a 
broadband spectroscopic analysis of the event using the WFC and GRBM data.  
However, their analysis did not yield sufficiently constrained spectral 
parameters, due to the much lower sensitivity and poorer energy resolution of 
GRBM compared to those of the LADs. We performed here a broadband time-resolved
spectral analysis of GRB\,980425, combining the LAD and the WFC data, to 
better constrain the spectral parameters and their evolution during the burst.

We used the High Energy Resolution Burst data of the brightest LAD for this 
event, with 128-channel energy resolution \citep{kan06}. This datatype starts 
after the BATSE trigger time, which corresponds to the beginning of interval A, 
as shown in Figure~\ref{fig:980425_lc}. We modeled the LAD background with a 
low-order polynomial function on spectra taken $\sim$~800 s before and 
$\sim$~200 s after the burst episode. The WFC-2 detected the burst at a fairly 
large off-axis angle of 18\fdg 7 with a light collecting area of only 10\,cm$^2$ 
(6\% of the on-axis value). However, since the WFC is an imaging telescope with 
a coded mask, all imaged data can be corrected for background without ambiguity.
In addition to the background correction, the data were corrected for the 
detector deadtime as well as for the variations of the spectral response over 
the detector area.

To assure good statistics, we binned both the LAD and WFC data into four 
time intervals (A, B, C, and D in Figure~\ref{fig:980425_lc}); two intervals 
each before and after the WFC lightcurve peak, similar to the ones used by 
\citet{fro00}. We analyzed the data from both detectors jointly, using the 
spectral analysis software RMFIT \citep{rmfit}. Since no significant signal was 
found above 300\,keV in all intervals ($\leq 2.2 \sigma$ detection in each 
interval), we obtained the best fits using the ``Comptonization'' photon model
\citep{kan06}:
\begin{displaymath}
f({\mathcal E}) = 
A \left(\frac{\mathcal E}{100\;\rm{keV}}\right)^{\textstyle \alpha}
\exp\left[-\frac{(2 + \alpha){\mathcal E}}{\epeak}\right]. \\
\end{displaymath}
Here, $A$ is the amplitude in photons$\;{\rm s^{-1}\;cm^{-2}\;keV^{-1}}$, 
$\alpha$ is the power-law index, and $\epeak$ is the photon energy 
(${\mathcal E}$) where $\nu F_{\nu} \propto {\mathcal E}^2f({\mathcal E})$ peaks. 
We found no indication that interstellar absorption was required in the fit 
and the absorption was, therefore, not used. Finally, no normalization 
factor was needed between the datasets of the two detectors.

\begin{figure}
\centerline{
\plotone{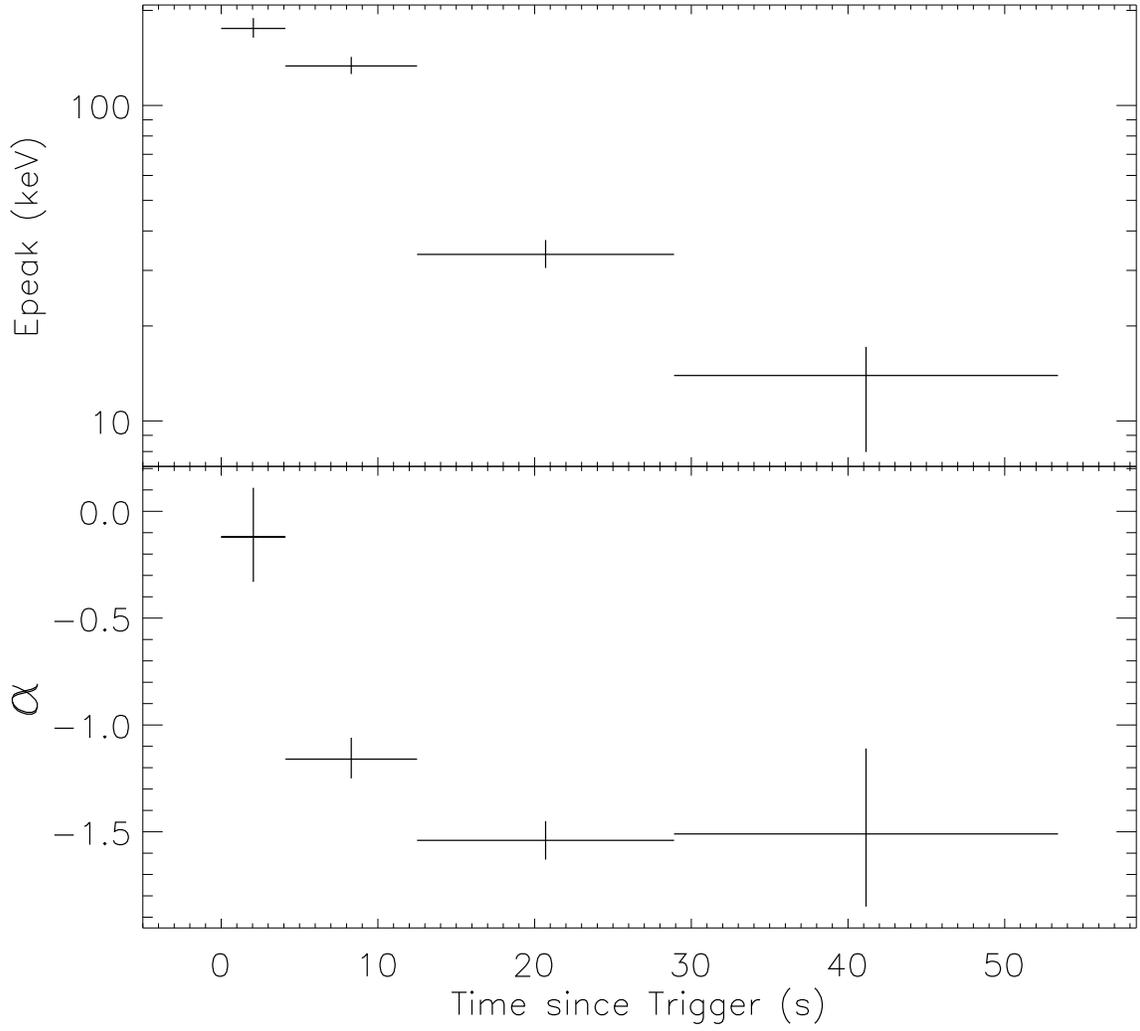}}
\caption{Spectral parameter evolution of GRB\,980425; the data 
points correspond to the intervals indicated in Figure~\ref{fig:980425_lc}.
The uncertainties are 1$\sigma$.}
\label{fig:980425_par}
\end{figure}

The spectral fitting results are summarized in Table~\ref{tab:980425_results}.  
Figure~\ref{fig:980425_par} displays the evolution of $\epeak$ and $\alpha$; 
both parameters clearly change from hard to soft, in line with the common GRB 
trend \citep[e.g.,][]{for95, cri97}. Moreover, the hardest part of the burst 
has an $\epeak$ of $175\pm13$\,keV, which is within 1$\sigma$ of the peak of 
the $\epeak$ distribution for time-resolved BATSE GRB spectra \citep{kan06}. 
Further, we found the $\epeak$ of the duration-integrated spectrum to be
$122 \pm 17$\,keV (1$\sigma$ uncertainty), placing the event within 2$\sigma$ 
from the peak of the distribution of BATSE GRBs \citep{kan06}. We find 
an isotropic-equivalent total energy emitted in 1$-$10000\,keV of $\egiso = 
(9.29 \pm 0.35) \times 10^{47}$\,erg, using a luminosity distance of 
$d_L = 35.6\;$Mpc \citep[for $z = 0.0085$;][]{tin98}.

The total energy fluence values in the X-ray and $\gamma$-ray bands
are $S_{X\;(2-30\;{\rm keV})} = 1.99 \times 10^{-6}$\,erg~cm$^{-2}$ and
$S_{\gamma\;(30-400\;{\rm keV})} = 3.40 \times 10^{-6}$\,erg~cm$^{-2}$,
respectively.  The ratio of the two, $\log(S_{X}/S_{\gamma}) = -0.23$,
makes GRB\,980425 an X-ray rich GRB, following the definition of \citet{lam04}
and \citet{sak05}.

\subsection{GRB\,030329}
\begin{figure}
\centerline{
\plotone{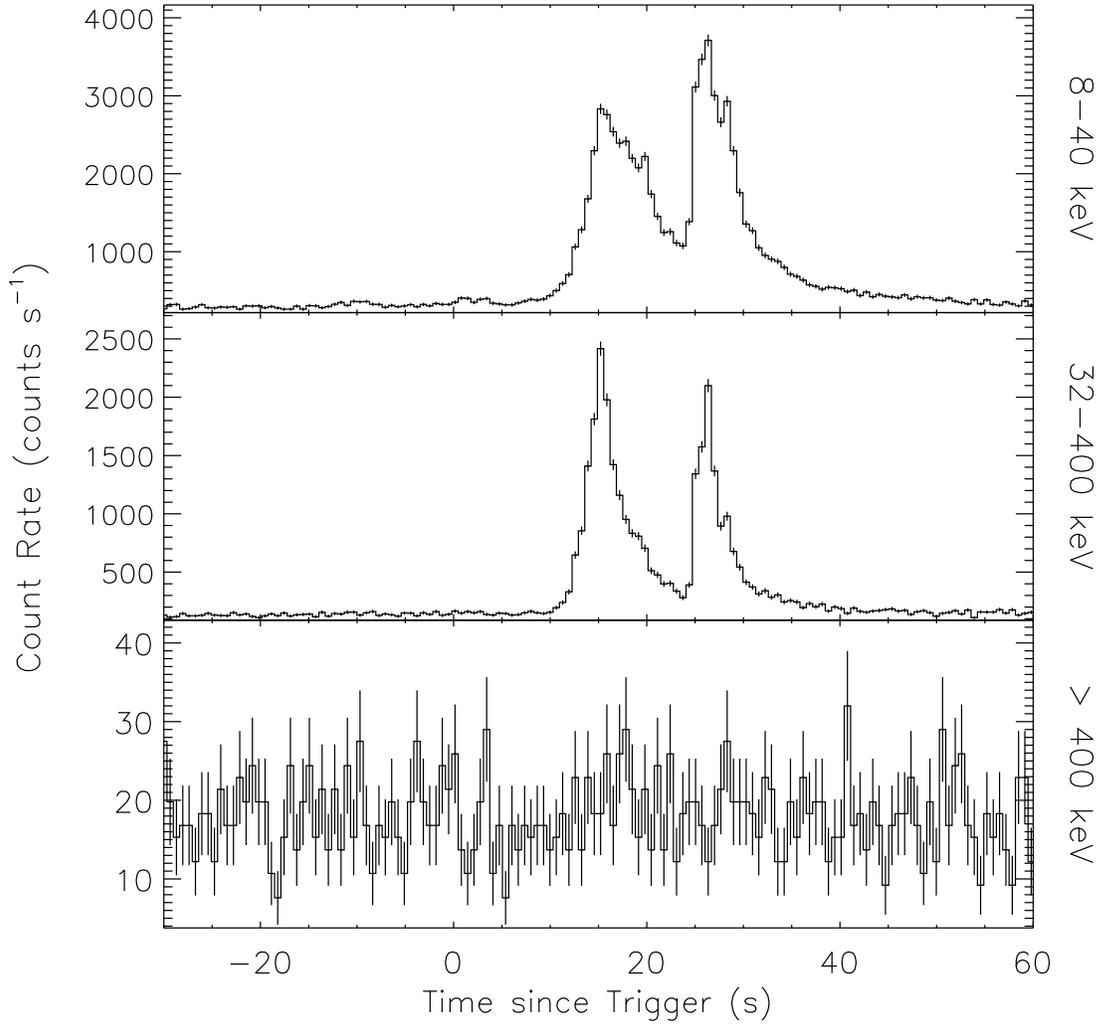}}
\caption{The {\it HETE-2}/FREGATE lightcurve of GRB\,030329 generated using 
the publicly available Burst Lightcurve data integrated in 64-ms resolution 
time bins.}
\label{fig:030329_lc}
\end{figure}

GRB\,030329 was detected with the FREGATE detector (8$-$400\,keV) on board the 
{\it High-Energy Transient Explorer (HETE-2)} on 2003 March 29 at 11:37:14 UT 
\citep[also Figure~\ref{fig:030329_lc}]{van04}. A detailed spectral analysis 
of the event was performed by \citet{van04}, who reported an $\epeak = 70.2 
\pm 2.3$\,keV for the duration-integrated spectrum. Based on their spectral
parameters and flux values, we estimate $\egiso = 1.33 \times 10^{52}\;$erg 
(1$-$10000\,keV) using $d_L = 790.8\;$Mpc ($z = 0.1685$; \citealt{Gre03}; 
see also Table \ref{tab:comparison}).

\subsection{GRB\,031203}
\begin{figure}
\centerline{
\plotone{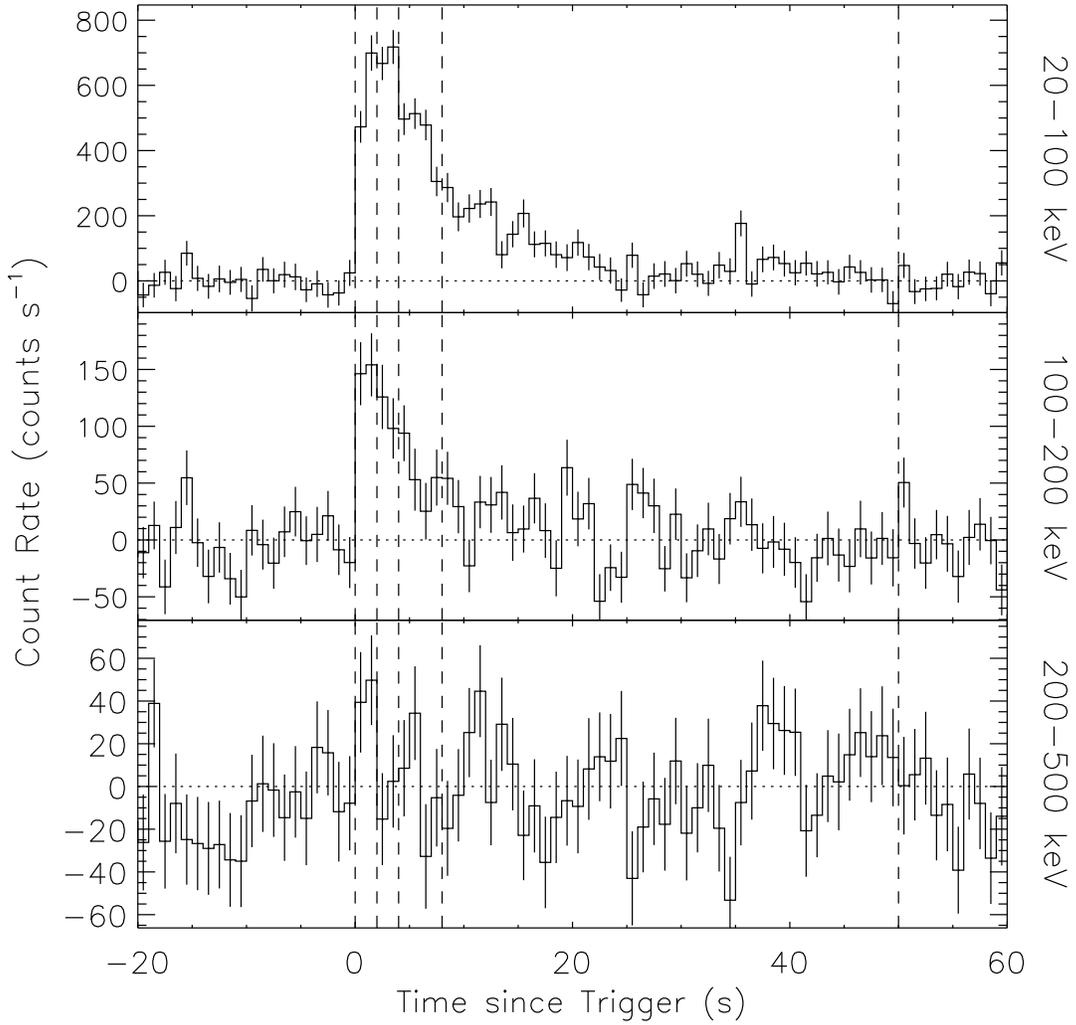}}
\caption{The {\it INTEGRAL} IBIS/ISGRI lightcurve of GRB\,031203 in 1-s 
resolution. The four time intervals used in the current analysis are indicated 
with the dashed lines. The dotted lines show the background levels.}
\label{fig:031203_lc}
\end{figure}

GRB\,031203 was detected with the IBIS/ISGRI detector (15$-$500\,keV) on the
{\it INTEGRAL} satellite on 2003 December 3 at 22:01:28 UT \citep{got03}.  
We obtained the IBIS/ISGRI data from the INTEGRAL Science Data Center, and 
processed them with OSA 5.0. Figure~\ref{fig:031203_lc} displays the lightcurve 
of this event, for which we estimated a $\gdur$ duration of 37.0\e1.3\,s. We 
extracted a background spectrum at $\sim$~300 seconds prior to the burst and a 
source spectrum encompassing the $\gdur \pm 10$\,s. In addition, we extracted 
four time-resolved spectra, with durations determined by requiring their 
signal-to-noise ratios to be above 35, for sufficient statistics. The four 
intervals are indicated in Figure~\ref{fig:031203_lc}.

We used XSPEC v12.2.1 for all spectral analyses of this event. The extracted 
integral source spectrum (17$-$500\,keV; effective exposure time of 45.9 s) 
is best described by a single power law of index of $-1.71 \pm 0.08$, consistent 
with the results of \citet{saz04}. To constrain the $\epeak$, \citet{saz04} 
fitted the spectrum with the Band function \citep[two power laws smoothly 
joined at a break energy;][]{ban93} with fixed high-energy photon index, and
derived $\epeak > 190$\,keV (90\% confidence level). Employing the same method, 
we found $\epeak > 71\;$keV ($1\sigma$) and 36\,keV (90\%), and a low-energy 
photon index of $\alpha = -1.39 \pm 0.41$. These results are consistent with 
\citet{saz04} since the longer integration time used here (45.9\,s {\it versus} 
22\,s) includes the softer tail portion of the event, while the background 
contribution is practically negligible. Using the single power-law fit 
parameters we estimate $S_{X}/S_{\gamma} = 0.49$ (Table~\ref{tab:comparison}), 
which is in the middle of the range predicted by \citet{saz04}. Therefore, 
although the {\it INTEGRAL} data alone cannot confirm an X-Ray-Flash (XRF) nature 
for GRB\,031203, as suggested by \citet{wat04}, we find the event to be clearly 
X-ray rich.

\begin{figure}
\centerline{
\plotone{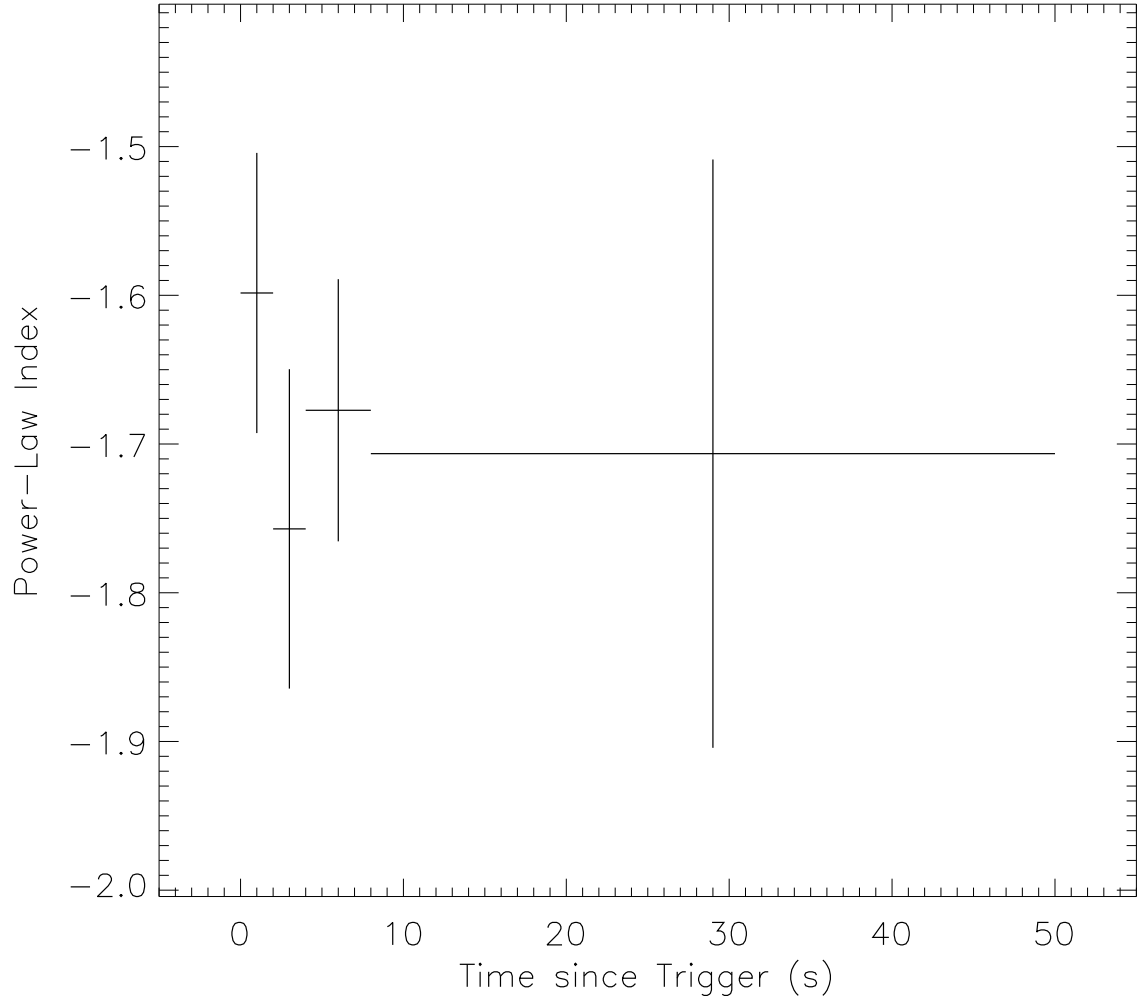}}
\caption{Single power law spectral index evolution of GRB\,031203; the data 
points correspond to the intervals indicated in Figure~\ref{fig:031203_lc}.}
\label{fig:031203_par}
\end{figure}

The isotropic-equivalent total emitted energy was estimated extrapolating the
single power law fit between 1$-$10000\,keV in the source rest frame
\citep[$z = 0.105$;][]{pro04} leading to an $\egiso = (1.67^{+0.04}_{-0.10}) 
\times 10^{50}\;$erg, using $d_L = 472.6\;$Mpc.  However, since we assumed 
no spectral break up to 10 MeV, we consider this estimate to be an upper limit 
of $\egiso$.  All time-resolved spectra were also best fitted with a single 
power-law model; the fit results are shown in Table~\ref{tab:031203_results} 
and the power-law index evolution is presented in Figure~\ref{fig:031203_par}. 
In accord with \citet{saz04} and as illustrated in Figure~\ref{fig:031203_par}, 
we found no significant spectral evolution within the burst, 
although spectral softening is suggested by the lightcurve of this event
(Figure~\ref{fig:031203_lc}; see also \S \ref{sec:spec_evol}).

\subsection{GRB\,060218}

\begin{figure}
\centerline{
\plotone{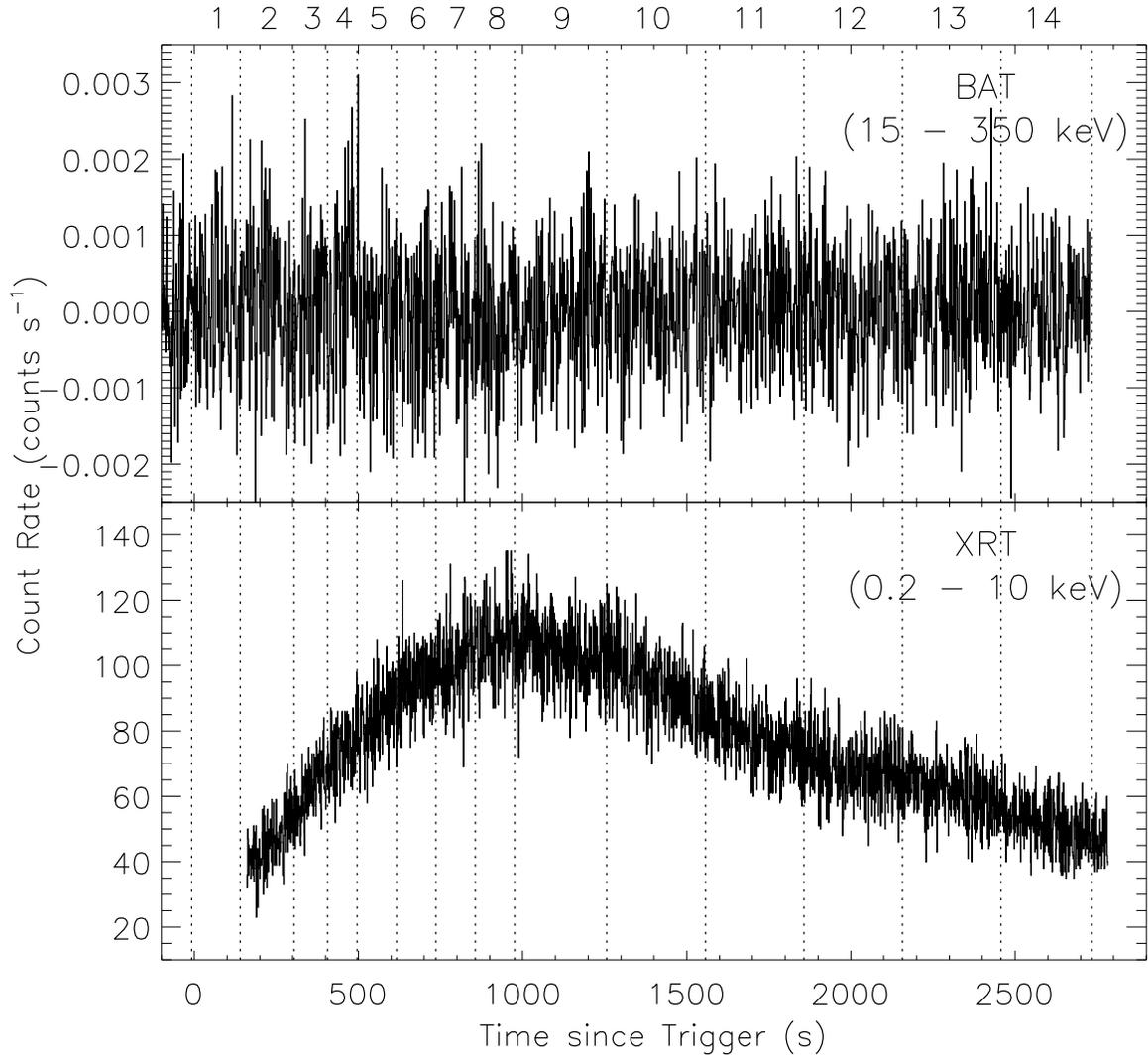}}
\caption{The lightcurve of GRB\,060218 seen with {\it Swift} BAT (15$-$350\,keV;
{\it top}) and XRT (0.2$-$10\,keV; {\it bottom}); the data are binned with 
1.6 and 1.0\,s resolution, respectively. The 14 time intervals used in our 
time-resolved analysis are indicated with dotted lines.}
\label{fig:060218_lc}
\end{figure}

GRB\,060218 was detected with the Burst Alert Telescope (BAT; 15$-$350\,keV) 
on board the {\it Swift} satellite, on 2006 February 18, at 3:34:30 UT 
\citep{cus06}. The burst was exceptionally long, with a $\gdur$ duration of 
$2100 \pm 100\;$s \citep{cam06}. The available BAT data types were Event data 
up to $\sim$~300\,s after the trigger time (T$_0$) and Detector Plane Histogram 
(DPH) data, thereafter. The latter data type has a variable time resolution; 
when in survey mode, the nominal resolution is 300\,s, but shorter exposures 
(with a minimum of 64\,s) are available immediately after a burst 
trigger\footnote{BAT Data Products v3.1 by H. Krimm
(\url{http://www.swift.ac.uk/BAT\_Data\_Products\_v3.1.pdf}).}. After the
burst trigger, the satellite slewed immediately to the BAT burst location and 
the X-Ray Telescope (XRT; 0.2$-$10\,keV) started observing the burst around 
T$_0 + 160\;$s. The XRT data were in Windowed Timing (WT) mode from 
161 to 2770\,s after T$_0$. Figure~\ref{fig:060218_lc} shows the BAT and XRT WT 
lightcurves of GRB\,060218.  
Notice the much higher photon flux in the X-ray lightcurve, which clearly 
indicates the XRF nature of this event.
In fact, the burst was detected with BAT by an image trigger, which typically 
displays no significant signals in the BAT lightcurve \citep{cus06}.

We downloaded the publicly available BAT and XRT data and processed them with 
FTOOLS v6.0.4. For the present analysis, we used the BAT Event data for the 
first 150 seconds of the burst, and the BAT DPH and XRT WT data for the rest 
of the burst. This combination of burst trigger and survey mode led to 14 time 
bins. The time bins are also shown in Figure~\ref{fig:060218_lc}. Before 
extracting spectra, we rebinned the DPH data with {\tt baterebin} to the 
correct energy calibration edges. Then we created the mask-weighting map 
for the burst location using {\tt batmaskwtimg} to extract the mask-weighted 
spectra from the rebinned DPH data. Finally, we extracted the spectra using 
{\tt batbinevt} for both the Event and DPH data, and generated the 
corresponding detector response matrices with {\tt batdrmgen}. For the XRT 
data, we obtained the screened WT event file and extracted spectra for the 13 
different time intervals corresponding to the DPH time intervals. We used 
{\tt xselect} and an extraction region of 50\arcsec. For each spectrum, we 
generated the corresponding Auxiliary Response Files (ARF) with {\tt xrtmkarf} 
using the default empirical ARF file as an input. The spectra were then grouped 
so that each energy channel contained a minimum of 20 counts, and the latest
available response matrix (v007) was used for the analysis. No pile-up
corrections were necessary for this event.

We analyzed with XSPEC v12.2.1 the BAT and XRT data jointly, except for the 
first time interval for which only BAT Event data were available. The BAT and 
XRT energy ranges used for the analysis were 15$-$150\,keV and 0.6$-$9\,keV, 
respectively. In the majority of these intervals, an absorbed power law with 
an exponential high-energy cutoff was the best fit, albeit with a significant 
low energy excess. This excess was also identified by \citet{cam06} and was 
fitted with an additional blackbody (BB) component. We have, therefore, 
included the BB in all spectral fits where the XRT data were included.

The BAT Event data (T$_0-8$ to T$_0+140$) spectrum is best fitted with a single 
power law with a high-energy cutoff, with spectral index of $-0.87 \pm 0.75$ 
and $\epeak$ of $24.9 \pm 6.0$\,keV (see also Table~\ref{tab:060218_results}).  
The estimated energy fluence during this interval (0.5$-$150\,keV) is 
$(6.38 \pm 4.00) \times 10^{-7}$\,erg\,cm$^{-2}$. From T$_0+140$ to T$_0+2734$, 
we analyzed the BAT-XRT joint spectrum; the best fit was an absorbed Band 
function with an additional BB component of $kT = 0.150 \pm 0.004\;$keV. The 
low and high energy spectral index and the peak photon energy values were 
$\alpha = -1.44 \pm 0.06$, $\beta = -2.54 \pm 0.07$, and $\epeak = 4.67 \pm 
1.15$\,keV, respectively. The absorption model used was {\tt wabs}, with the 
best-fit value of $N_H = (0.60 \pm 0.02) \times 10^{22}$\,cm$^{-2}$. The 
unabsorbed energy fluence (0.5$-$150\,keV) was $(1.72^{~+0.18}_{~-0.78}) \times 
10^{-5}$\,erg\,cm$^{-2}$, where the errors were estimated from the absorbed 
flux. We derived an isotropic-equivalent total emitted energy of $\egiso =
(4.33^{~+0.41}_{~-1.74}) \times 10^{49}\;$erg in 1$-$10000\,keV (source frame),
using $d_L = 143.2\;$Mpc \citep[$z = 0.0335$;][]{mir06,pia06}.

\begin{figure}
\centerline{
\plotone{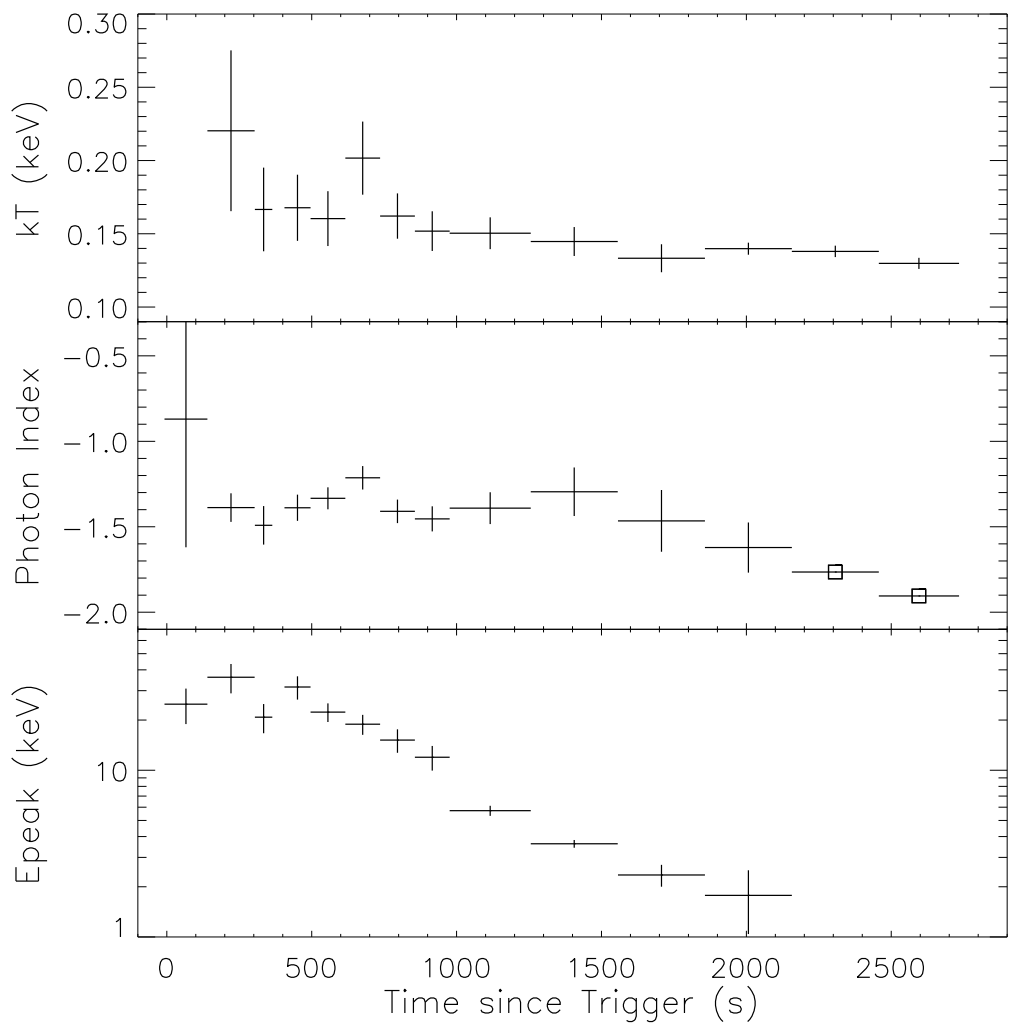}}
\caption{The spectral parameter evolution of GRB\,060218.  The last two
points in the photon index plot ({\it middle panel}, marked with
squares) were estimated by fitting a power law with exponential cutoff
with the exponential cutoff energy fixed to the last well-determined value
(4.7\,keV).  All the uncertainties are 1$\sigma$.}
\label{fig:060218_par}
\end{figure}

The results of the time-resolved spectral fits are listed in detail in 
Table~\ref{tab:060218_results}. We fitted an absorbed power law with a 
high-energy cutoff and a BB component in all BAT-XRT joint spectra but the last 
two, for which a single power law with BB provided adequate fits. The $N_H$ 
values were found to be constant ($N_H \sim 0.6 \times 10^{22}$\,cm$^{-2}$) 
throughout the first 10 BAT-XRT joint time intervals, and were fixed to that 
value for the last three spectra, where their best-fit $N_H$ values were higher 
resulting in excessive values of unabsorbed flux. Fixing $N_H$ did not affect 
significantly the other parameters in the fits. Figure~\ref{fig:060218_par} 
displays the spectral parameter evolution of the 14 time-resolved spectra. We 
observed a hard-to-soft spectral evolution in the non-thermal spectra, while 
the BB temperature ($kT$) remained constant. On the average, the BB component 
contributes 0.13\% of the source flux throughout the burst prompt emission 
duration.

\subsection{Comparison}

\begin{figure}
\centerline{
\plotone{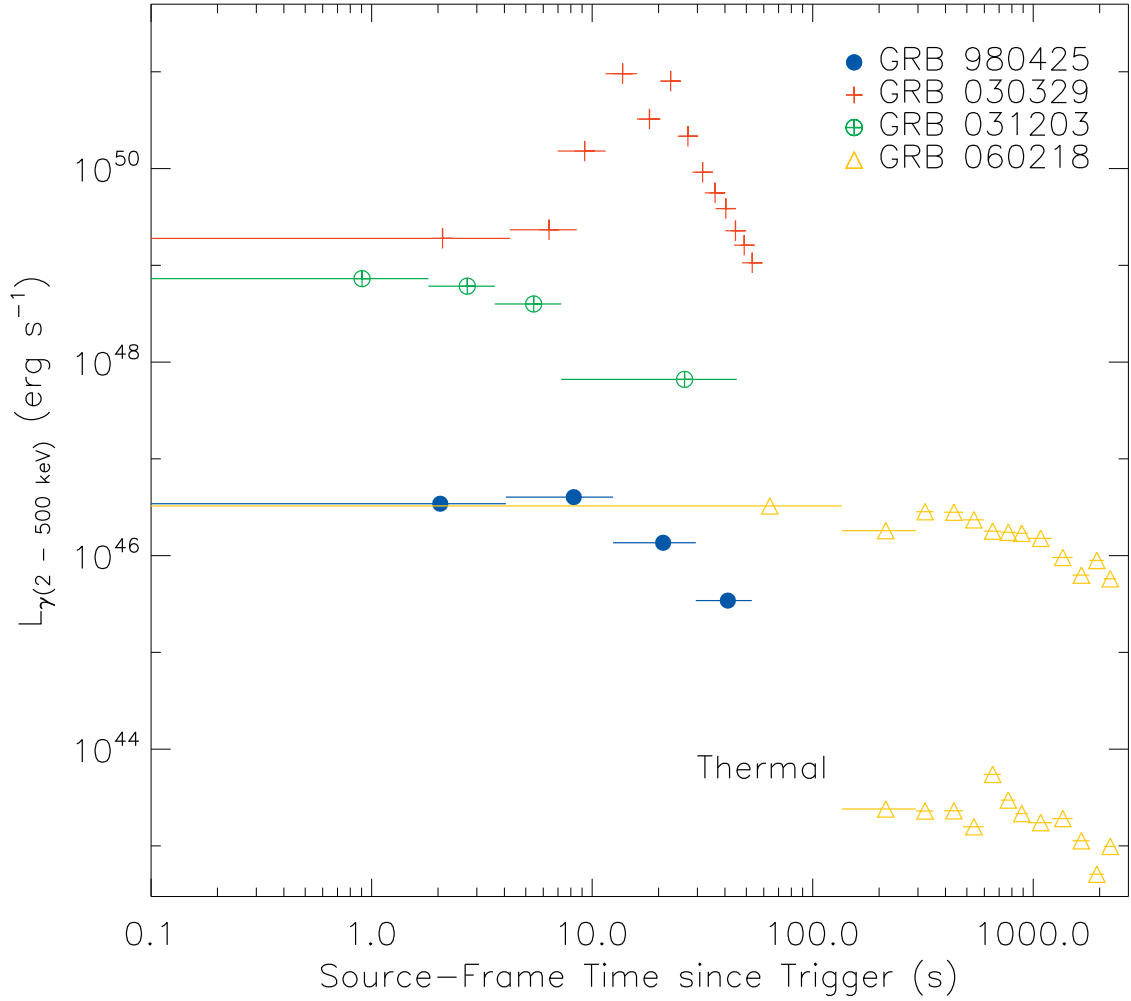}}
\caption{The evolution of the $\gamma$-ray isotropic-equivalent luminosity 
(2$-$500\,keV, source-frame energy) for all four events, in the source-frame 
time. For GRB\,060218, the luminosity of the thermal component is plotted 
separately.}
\label{fig:lgamma}
\end{figure}

The $\gamma$-ray properties of the four events are summarized in 
Table~\ref{tab:comparison}, and Figure~\ref{fig:lgamma} shows the evolution
of their isotropic-equivalent $\gamma$-ray luminosity, $\lgiso$ (in 
2$-$500\,keV, source-frame energy). For GRB\,060218, the contributions from the 
thermal component are shown separately from the $\lgiso$ of the non-thermal 
component. We find that GRB\,060218 had similar $\gamma$-ray luminosity as 
GRB\,980425; both were a few orders of magnitude fainter than the other two 
events. However, since GRB\,060218 lasted considerably longer than GRB\,980425, 
its $\egiso$ was $\sim 50$ times larger.

We should emphasize here that we measure directly only the energy radiated in 
the direction of the Earth per second per steradian per logarithmic energy 
interval by a source at luminosity distance $d_L$. The apparent bolometric 
luminosity may be quite different from the {\it true} bolometric luminosity if 
the source is not isotropic.

To compare the lightcurves of these events, we fitted a two-sided Gaussian 
function to the main pulses of the lightcurves. We found that the time profiles 
of these events are similar to the overall profile found for GRBs \citep{nem94},
with the rising part having a Half Width Half Maximum (HWHM) much smaller 
($< 1/2$) than the HWHM of the decaying part.

\subsubsection{Duration-Integrated Spectra} 
\label{sec:comp1}

\begin{figure}
\centerline{
\plotone{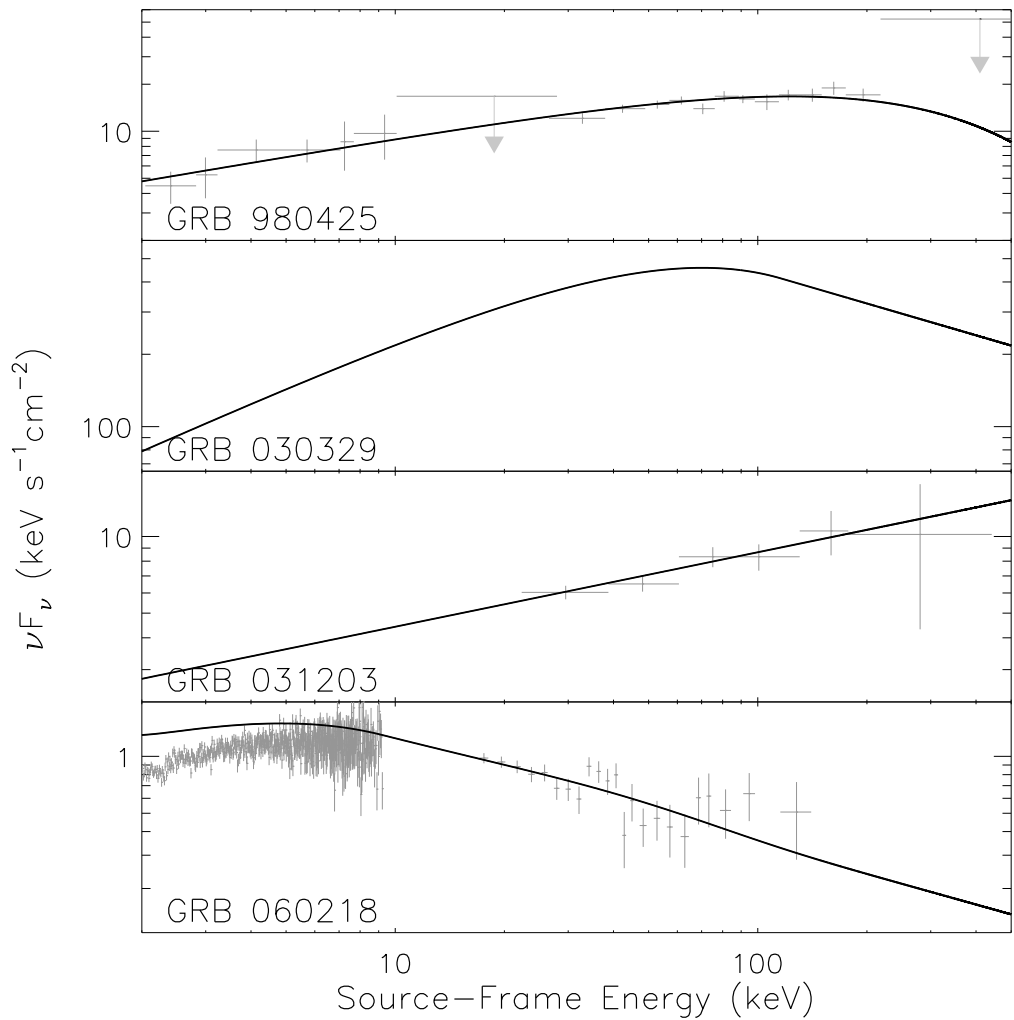}}
\caption{The unabsorbed best-fit duration-integrated spectra (solid lines) of 
the four events, overplotted with the deconvolved data (gray crosses) in the 
source-frame energy. The data are binned for display purposes.
The analysis tool for the {\it HETE-2} data of GRB\,030329 was not publicaly
available, and we only show here the spectral model presented in \citet{van04}.
See also text in \S~\ref{sec:comp1}.}
\label{fig:spectra}
\end{figure}

Figure~\ref{fig:spectra} compares the best-fit duration-integrated spectra 
($2-500$\,keV in source frame) of the four events in $\nu F_\nu$, which shows 
the energy radiated per logarithmic photon energy interval. We plot here the 
unabsorbed spectral models in each case (solid lines) together with the 
spectrally deconvolved, blue-shifted data points shown in gray. The data are 
binned here for display purposes. No absorption correction was necessary in 
the combined LAD-WFC fit of GRB\,980425. We used the published spectrum for 
GRB\,030329, and we extrapolated our best-fit model for GRB\,031203 to the 
lower energy range shown here. For GRB\,060218 the absorption correction was
significant as seen from the comparison of the model to the XRT data
below 10\,keV.

\begin{figure}
\centerline{
\plotone{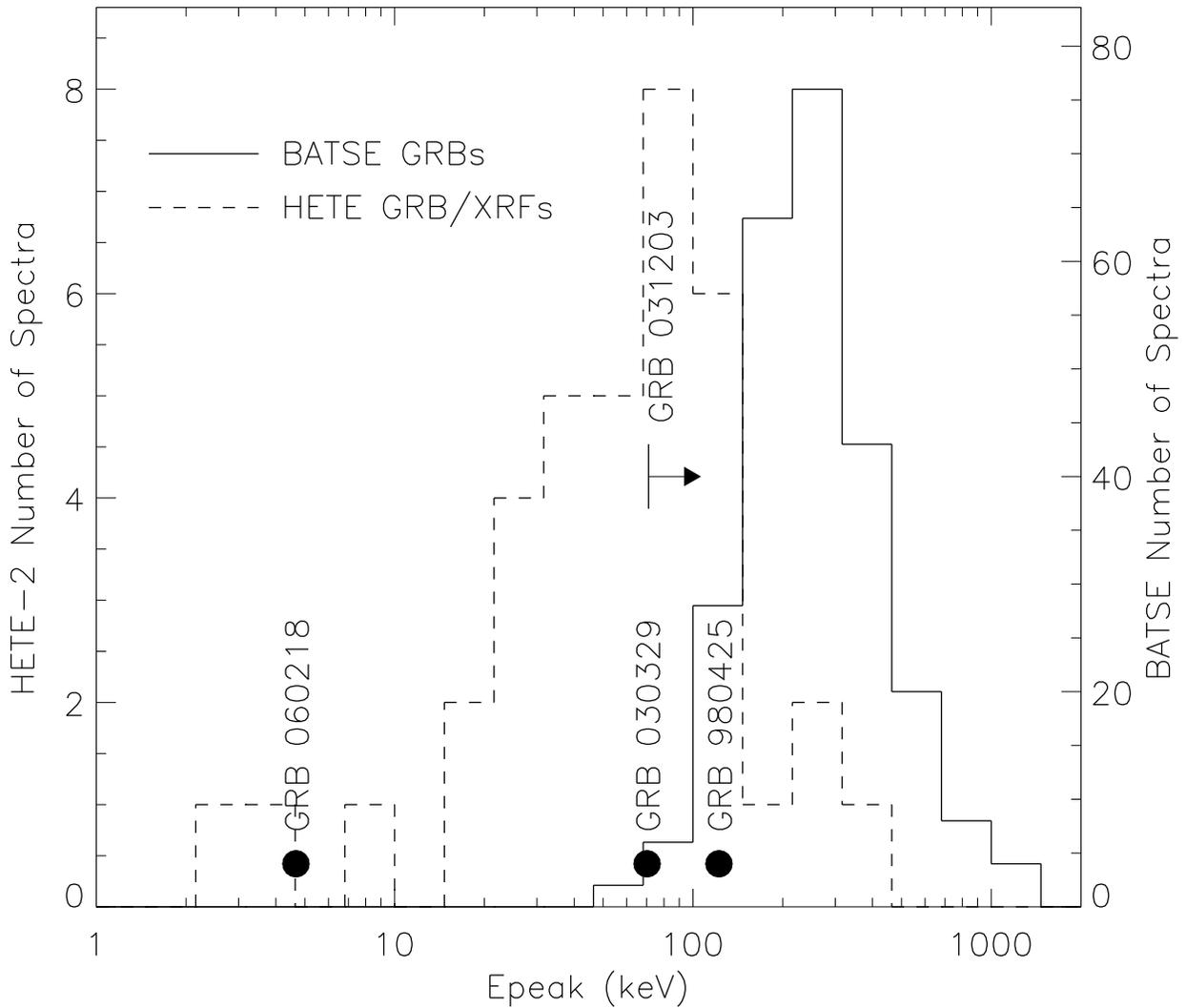}}
\caption{The $\epeak$ of the duration-integrated spectra of GRBs~980425, 
030329, and GRB\,060218; for GRB\,031203, we plot a 1$\sigma$ lower limit. As a 
comparison, the $\epeak$ distributions of 251 bright BATSE GRBs \citep{kan06} 
and 37 {\it HETE-2} GRB/XRFs \citep{sak05} are plotted.
Only well-constrained $\epeak$ values are included in the distributions.}
\label{fig:ep_dist}
\end{figure}

In Figure \ref{fig:ep_dist} we compare the $\nu F_\nu$ peak energy, $\epeak$, 
of the same duration-integrated spectra, to the $\epeak$ distributions of 251 
bright BATSE GRBs \citep{kan06} and 37 {\it HETE-2} GRB/XRFs \citep{sak05}. The 
$\epeak$ values of the four events are lower (softer) than the average value of 
the bright BATSE GRBs but they all seem to fit well within the {\it HETE-2} 
$\epeak$ distribution. Note that the BATSE $\epeak$ distribution shown here is 
derived using the brightest BATSE GRBs, which tend to be spectrally harder than 
dim GRBs \citep{mal95}.

\begin{figure}
\centerline{
\plotone{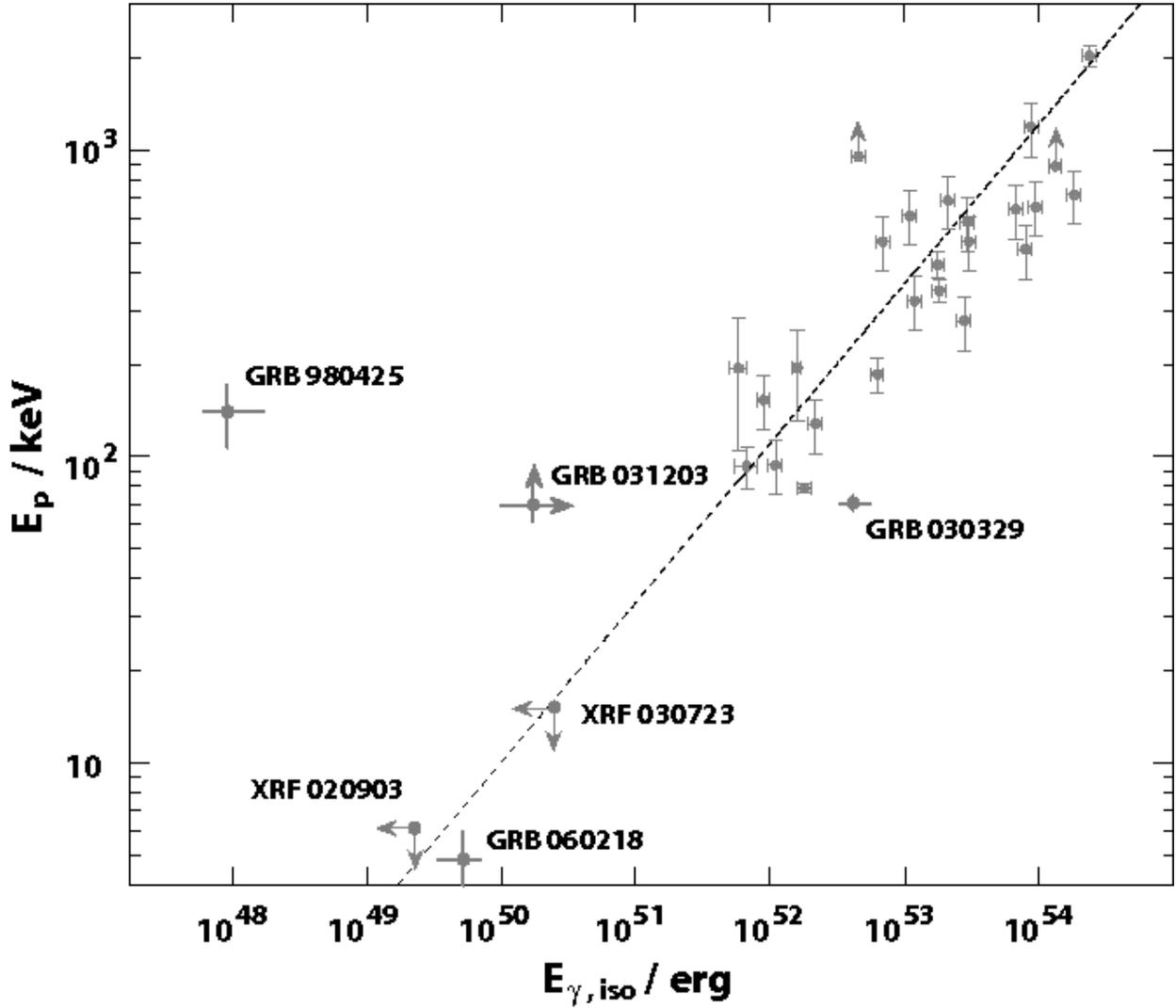}}
\caption{Locations of GRBs~980425, 031203, 030329, and 060218 in
the $\epeak$-$\egiso$ plane.  The dashed line indicates the correlation found 
by \citet{ama02}. The events presented in \citet{ghi04} are also shown here.}
\label{fig:ep_eiso}
\end{figure}

Finally, in Figure~\ref{fig:ep_eiso}, we show where these four events fall in 
the $\epeak$-$\egiso$ plane, in comparison to the empirical correlation (dashed 
line) found by \citet{ama02} and \citet{lrrr02}. GRB\,980425 lies farthest from 
the so-called Amati relation. Although GRB\,031203 also lies away from the 
relation, this deviation could be the result of a viewing angle slightly 
outside the edge of the jet \citep{ram05}. In fact, we find a smaller lower 
limit on $\epeak$ for this event than previously reported, so it is possible 
that, after correcting for an off-axis viewing angle, it would be consistent 
with the relation \citep[compare our Figure 11 to the left panel in Figure 3 
of][]{ram05}. 
We also note that for GRBs~980425 and 031203, \citet{ghi06} have recently 
suggested two other possible scenarios that may cause such a deviation from the
Amati relation: one is that the GRB radiation may have passed through a
scattering screen, and the other is that we may have missed softer emission from
the GRBs lasting much longer.  In both cases, observed spectra could have much
higher $\epeak$ values than the intrinsic ones.
Altogether, the four events do not appear to follow a specific 
pattern or correlation in the $\epeak$-$\egiso$ plane.

\subsubsection{Spectral Evolution}\label{sec:spec_evol}

\begin{figure}
\centerline{
\plotone{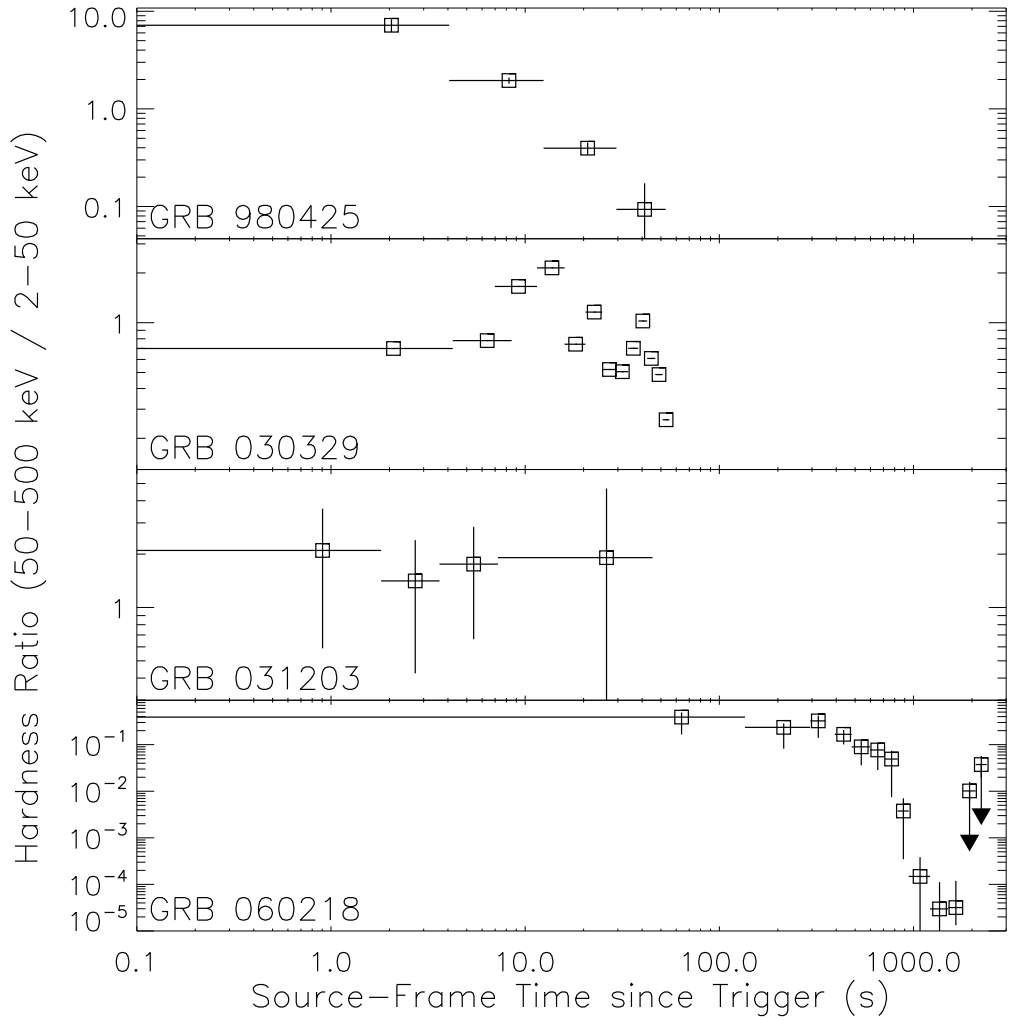}}
\caption{The hardness ratio evolution of all four events in the source-frame 
time. The hardness ratio is determined in the source-frame energy for each 
event. The values for GRB\,030329 were estimated from the spectral parameters 
presented in \citet{van04}.}
\label{fig:hardness}
\end{figure}

\begin{figure}
\centerline{
\plotone{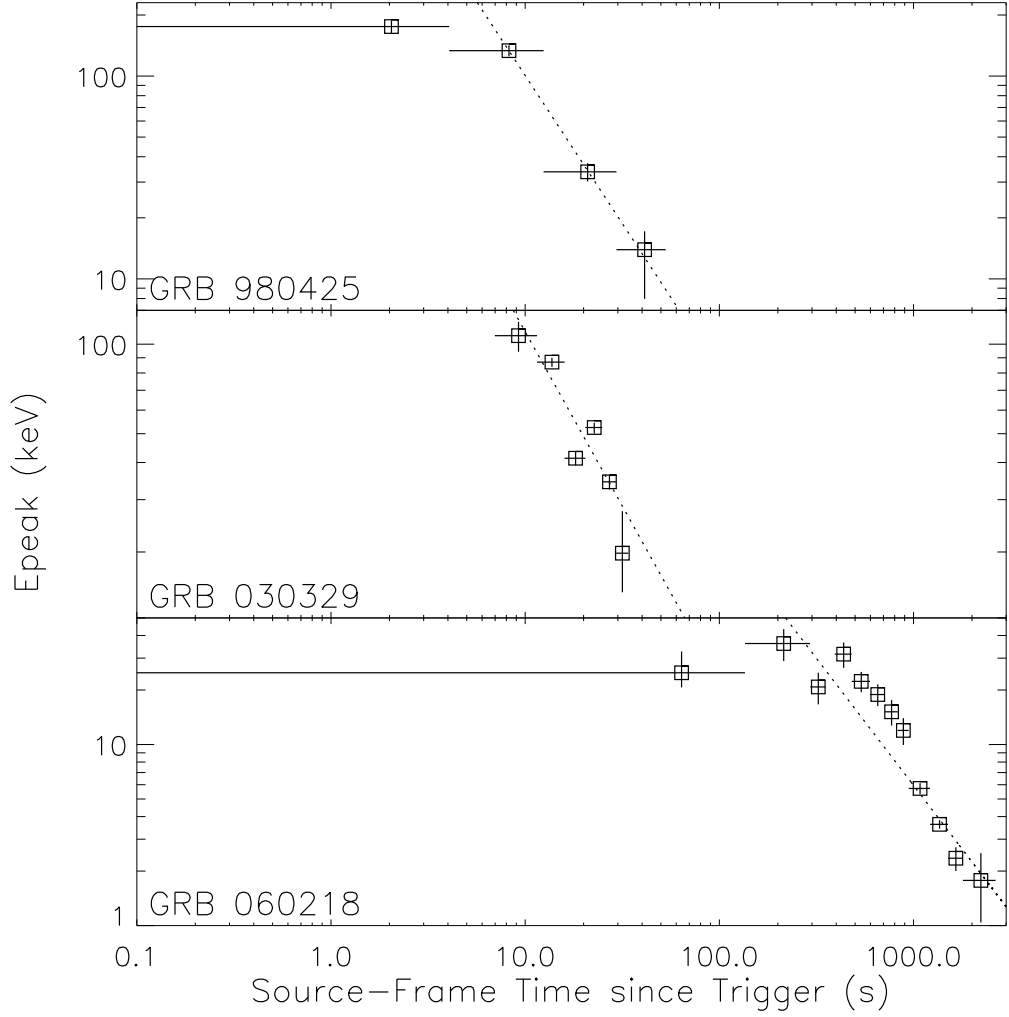}}
\caption{The evolution of the $\epeak$ in GRBs~980425, 030329, and 060218, 
in the source-frame time. The dotted lines show the best-fit power law decay, 
$\epeak \propto t^{\phi}$, with $\phi = -1.46 \pm 0.12$, $-1.17 \pm 0.08$, and
$-1.40 \pm 0.06$, respectively. For GRBs~980425 and 060218, the first points 
are excluded in the power-law fits.}
\label{fig:ep_evol}
\end{figure}

Figure~\ref{fig:hardness} shows the hardness ratio evolution of all four events,
in their source-frame time. The hardness ratio is defined here as the ratio of 
the energy flux between 50$-$500\,keV and 2$-$50\,keV in the source-frame 
energy. The hard-to-soft evolution is seen in all events except GRB\,031203. The 
lack of spectral evolution in GRB\,031203 is probably due to the fact that this 
burst was relatively dim and the observed energy range was fairly narrow. The 
burst was only detected below 200\,keV (see Figure~\ref{fig:031203_lc}) and the 
break energy seemed to always lie around or above this energy; therefore, the 
flux above this energy may have been overestimated. We also show in 
Figure~\ref{fig:ep_evol}, the evolution of the $\epeak$ in three events 
(GRBs~980425, 030329, and 060218) for which the $\epeak$ values were 
determined from their time-resolved spectra. To characterize the decaying 
behavior, we fitted a power law ($\epeak \propto t^{\phi}$) to each of the data 
sets. Since we are only interested in the decaying behavior, the first points 
of GRBs~980425 and 060218 were excluded from the fits. We found $\phi = 
-1.46 \pm 0.12$, $-1.17 \pm 0.08$, and $-1.40 \pm 0.06$, respectively; the 
best-fit power laws are shown as dotted lines in Figure \ref{fig:ep_evol}. 
We note that in the external shock model, for example, if $\epeak$ is 
identified with the typical synchrotron frequency $\nu_m$, it is expected to 
scale as $t^{-3/2}$ once most of the energy is transfered to the external 
medium and the self-similar deceleration phase sets in \citep{BM76}. This 
decaying index is actually similar to what we observed here, especially when 
taking into account that the asymptotic power law 
of $t^{-3/2}$ is expected to be approached gradually rather than immediately.

\section{Afterglow Emission}
\label{sec:AG}

\subsection{X-ray Afterglow}

\begin{figure}
\centerline{
\plotone{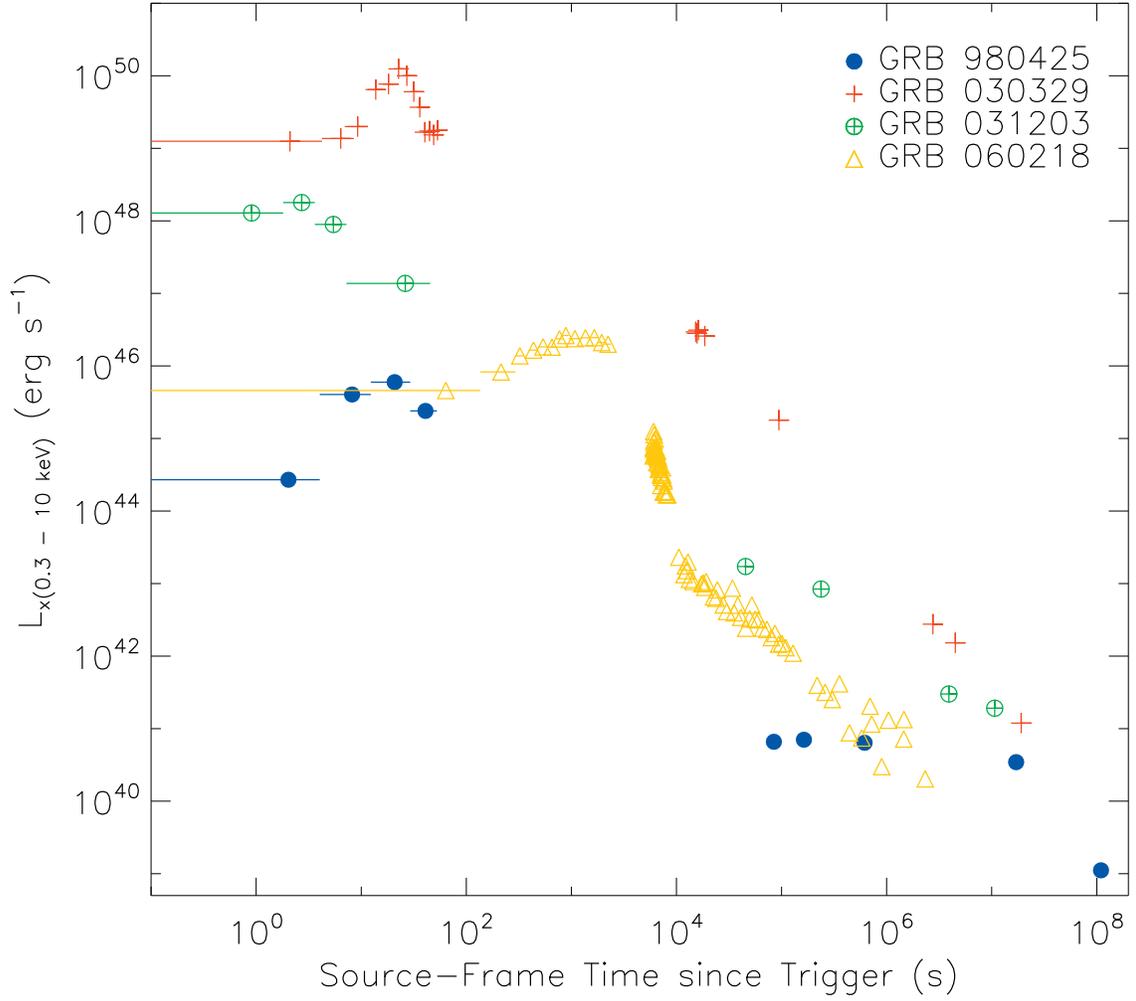}}
\caption{Evolution of the X-ray (0.3$-$10\,keV, source-frame energy)
prompt and afterglow luminosity (isotropic equivalent) in the source-frame time.}
\label{fig:lx_time}
\end{figure}

Figure~\ref{fig:lx_time} shows the evolution of the prompt and afterglow 
isotropic-equivalent luminosity in 0.3$-$10\,keV ($\lxiso$) in the source-frame 
time for all four events. We estimated the prompt X-ray luminosity values from 
the spectral analysis of the prompt $\gamma$-ray emission (\S \ref{sec:prompt}) by 
extrapolating the best-fit $\gamma$-ray spectra to the X-ray energy range for 
each event. The afterglow X-ray luminosity values were obtained in various ways.

For GRB\,030329, we derived the values using the data presented in 
\citet{tie03, tie04}. In the case of GRB\,031203, we analyzed the (now) archival 
{\it XMM} and {\it Chandra} observations. Our analysis results are consistent 
with those presented in \citet{wat04}. For GRB\,060218, we analyzed the 
{\it Swift} XRT photon counting (PC) mode data, starting $\sim 6200\;$s after 
the burst trigger. We extracted two time-averaged spectra, covering 
6191$-$8529\,s and $1.1 \times 10^4 - 1.3 \times 10^5$\,s after the burst 
trigger, respectively, to account for spectral evolution, and fitted each 
spectrum with a single power law. The spectral parameters were then used to 
convert the count rates to the luminosity values. The PC data after 
$1.3 \times 10^5$\,s were associated with relatively large uncertainties due to 
the fact that the count rates were approaching the XRT sensitivity limit, and 
therefore, we used the fitted parameters of the second spectrum above to 
estimate the luminosity values for these data points. We also include in this 
plot the two {\it Chandra} observations of GRB\,060218 at $\sim 10^6$\,s after 
the burst trigger, which were presented in \citet{sod06}.
Finally, we reanalyzed the {\it BeppoSAX} observations of GRB\,980425/SN\,1998bw, 
which were initially presented in \citet{pia00}. We used only the data from the 
{\it BeppoSAX}/Medium Energy Concentrator Spectrometers and followed closely 
the procedure described in \citet{pia00}. The only potentially significant 
differences between the current and the analysis of \citet{pia00} is that we 
estimated the background using the same observation, while \citet{pia00} used
calibration blank-sky observations. As a result, we found slightly lower net 
count rates (1.6$-$10.0\,keV) during the two observations in April 1998 
\citep[as compared to Table 1 in][]{pia00}, indicating a slightly flatter decay 
than was initially reported. In addition, we found that the observation 
beginning April 27 1998 contained $6.1\;$ks less data than previously reported 
(total of 14,862$\;$s exposure time). The count rates and exposure times for 
the remaining observations are consistent with \citet{pia00}.
 
\begin{figure}
\centerline{
\plotone{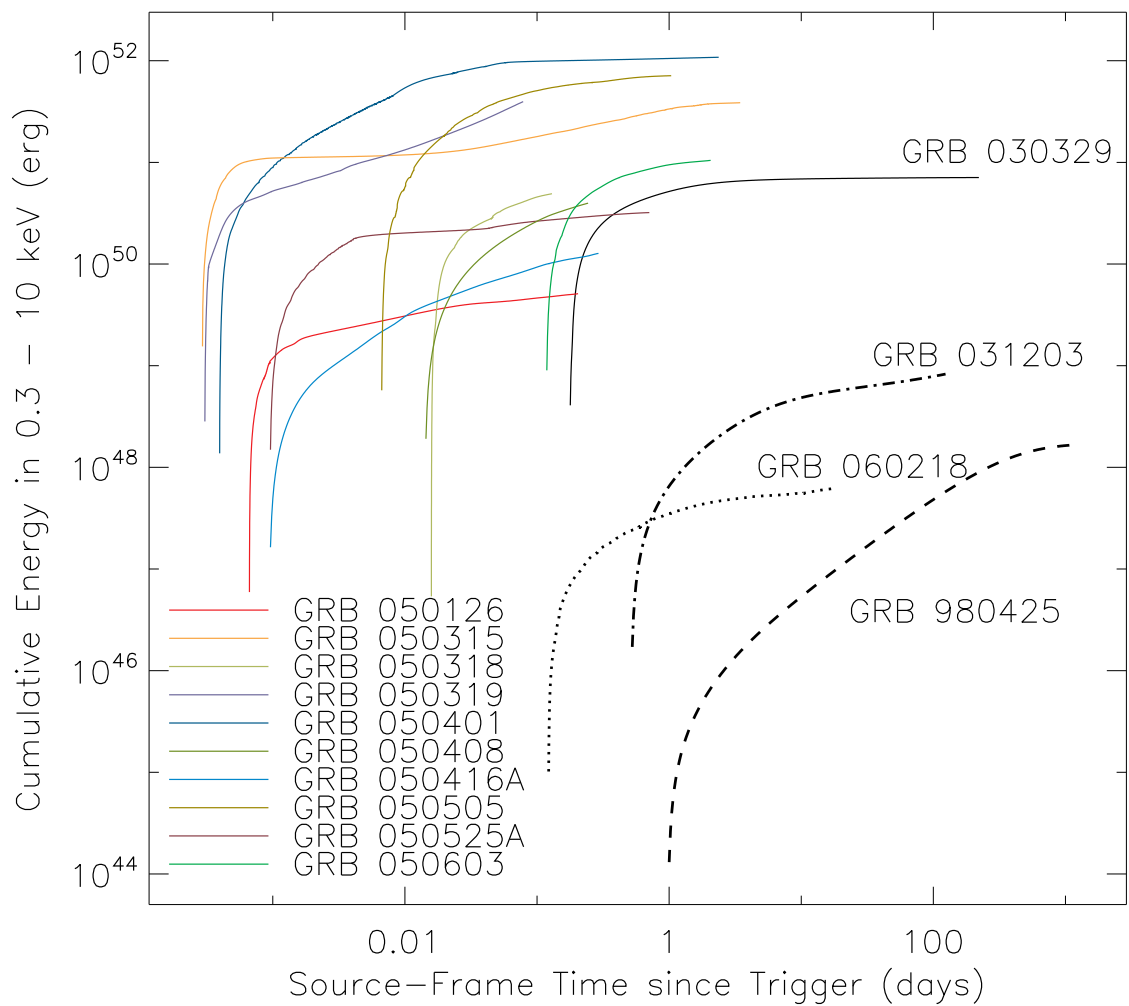}}
\caption{Cumulative total emitted energy (isotropic equivalent) of the four 
SN-GRBs (black lines) in 0.3$-$10\,keV (source-frame energy) as a function of 
source-frame time.
As a comparison, the cumulative energy of the 10 {\it Swift} GRBs with 
known redshifts published in \citet{nou06} are also shown here in color.}
\label{fig:lx_cum}
\end{figure}

To investigate the energy dissipation behavior in the X-ray afterglow, we fitted 
a natural cubic spline function to the $\lxiso$ history for each individual 
afterglow and estimated the cumulative emitted energy as a function of time. 
The evolution of the cumulative X-ray afterglow energy for all four events is 
shown in Figure~\ref{fig:lx_cum}. The integration time intervals varied from 
event to event, since the data time coverage was different for each event. We 
defined the integration time (i.e., ``afterglow'' time interval) of each event 
to be from the earliest X-ray observation time ($\sim 10^4$\,s) to the time of 
the latest available data point. In the case of GRB\,060218, however, it was 
difficult to determine the starting point of the afterglow since the prompt 
emission was extremely soft and long and the X-ray follow-up started much 
earlier than in the other three events. It is possible that the steep decay 
observed before $\sim 10^4$ s (see Figure~\ref{fig:lx_time}) belongs to the 
decaying part of the prompt emission; we, therefore, chose the afterglow 
starting time for this event to be $1.1 \times 10^4\;$s, which corresponds to 
the onset of the break in its $\lxiso$ history.

\begin{figure}
\centerline{
\plotone{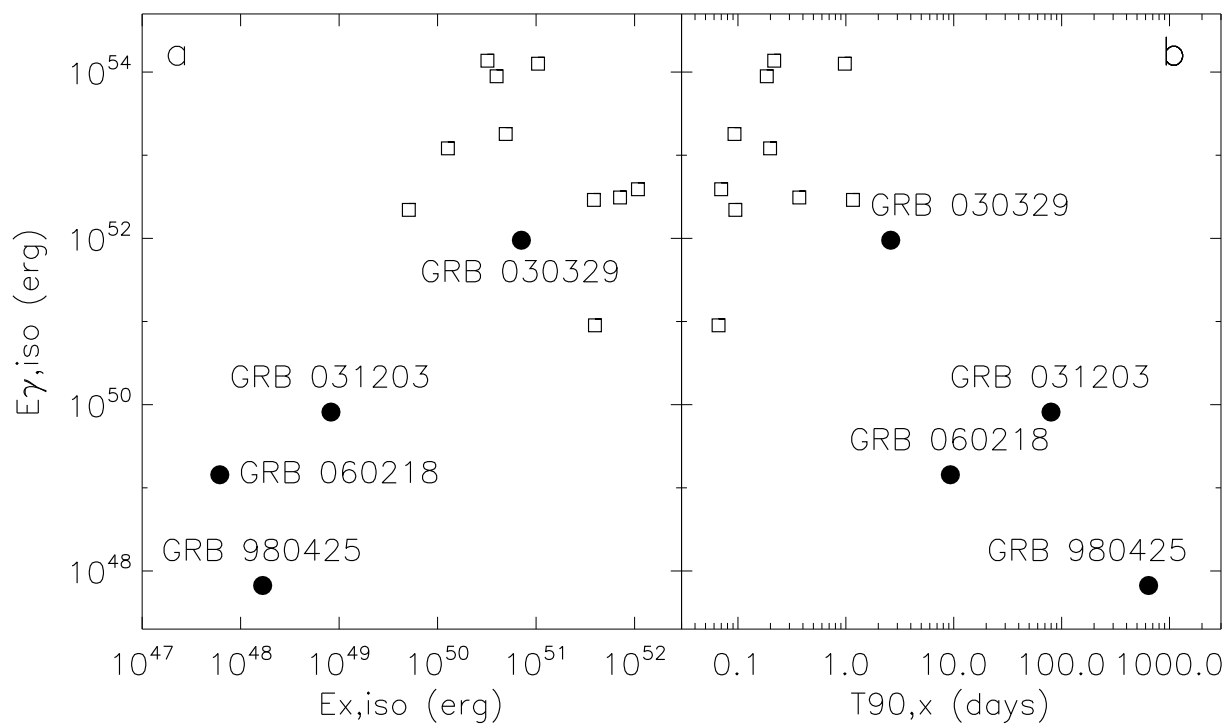}}
\caption{Comparison of the four SN-GRBs (full circles) with the 10 {\it Swift} 
GRBs presented by \citet{nou06} (squares), in the $\egiso$-$\exiso$ plane (a) 
and in the $\egiso$-$\xdur$ plane (b). The $\egiso$ values used here are 
determined between 20$-$2000\,keV in the source frame.}
\label{fig:egxisot90}
\end{figure}

To compare the cumulative energy evolution to regular GRBs, we also show in 
Figure~\ref{fig:lx_cum} the cumulative energy evolution of the 10 {\it Swift} 
GRBs with known redshifts that were presented in \citet{nou06}. For these 10 
events, we converted the $\lxiso$ values in 2$-$10\,keV presented there, to the 
$\lxiso$ in the energy range we used for our four events here (i.e., 
0.3$-$10\,keV), using the same spectral parameters used in \citet{nou06}, and 
again, fitted a spline function to each $\lxiso$ evolution. The start and end 
times of the integration were the first and the last points of the actual 
observations. Figure~\ref{fig:lx_cum} shows that the final $\exiso$ values for most of the 10 
\citet{nou06} GRBs span the range of $\sim 10^{50} - 10^{52}\;$erg. Out of the 
four SN-GRB events, only GRB\,030329 falls within this range, while the other 
three events fall between $\sim 10^{48}\;$ and $\sim 10^{49}\;$erg. 
We note that there exists a selection effect based on the observed photon flux:
an event would be more likely to be detected when it is closer to us than farther, 
for a given intrinsic luminosity. Therefore, the difference could be partially due 
to the fact that these four events occured relatively nearby compared to the 10 
{\it Swift} events. From the 
cumulative X-ray energy evolution, we also derived $\xdur$ values for all 
events, which we define here as the time in the source frame during which 90\% 
of the radiated afterglow energy (0.3$-$10\,keV in the source frame) is 
accumulated. Table~\ref{tab:comparison} displays the comparison of the X-ray 
afterglow properties of all four events, along with their $\gamma$-ray prompt 
properties. Again for the purpose of comparison, Figure~\ref{fig:egxisot90} 
shows the $\egiso$, $\exiso$, and $\xdur$ values of the four SN-GRBs, along 
with those of the 10 \citet{nou06} GRBs. Here we plot the $\egiso$ values 
between 20$-$2000\,keV (in the source frame). There is a clear correlation 
between $\egiso$ and $\exiso$ ($\egiso \propto \exiso^{1.0\pm 0.3}$), and an 
anti-correlation between $\egiso$ and $\xdur$ ($\egiso \propto \xdur^{-1.2\pm 
0.3}$). We also find $\xdur \propto \exiso^{-0.7\pm0.2}$, which can be 
interpreted as follows.

Events that have large isotropic-equivalent energy (both in $\gamma$-rays, 
$\egiso$, and in the kinetic energy of their afterglow, $\ekiso$) have a large 
$\exiso$, indicating a reasonably narrow spread in the efficiency of converting 
the afterglow kinetic energy into radiation. Moreover, they are typically 
associated with narrow jets. This means that most of their kinetic energy is in 
relativistic outflow carried by highly-relativistic ejecta 
\citep[with $\Gamma \sim 30-50$;][]{GK06}. Therefore, they have a relatively 
small $\xdur$; the time at which most of the energy is injected into the 
afterglow shock, of the order of $10^3\;{\rm s} \lesssim t_{\rm break,2} 
\lesssim 10^4\;$s \citep{nou06}. On the other hand, events that have a small 
$\ekiso$ and small $\egiso$, naturally have a small $\exiso$. Such events also 
tend to have most of their relativistic outflow energy residing in mildly 
relativistic ejecta, rather than in highly relativistic ejecta  
\citep{GR04,Waxman04}. As a result, most of the kinetic energy is transfered to 
the afterglow shock at relatively late times, hence a large $\xdur$. Further, 
since such events tend to be only mildly collimated, the {\it true} total 
energy in relativistic ($\Gamma \gtrsim 2$) ejecta has a significantly smaller 
spread than the one of the isotropic equivalent energy ($\egiso$ or $\exiso$).

\subsection{Radio Afterglow}

\begin{figure}
\centerline{
\plotone{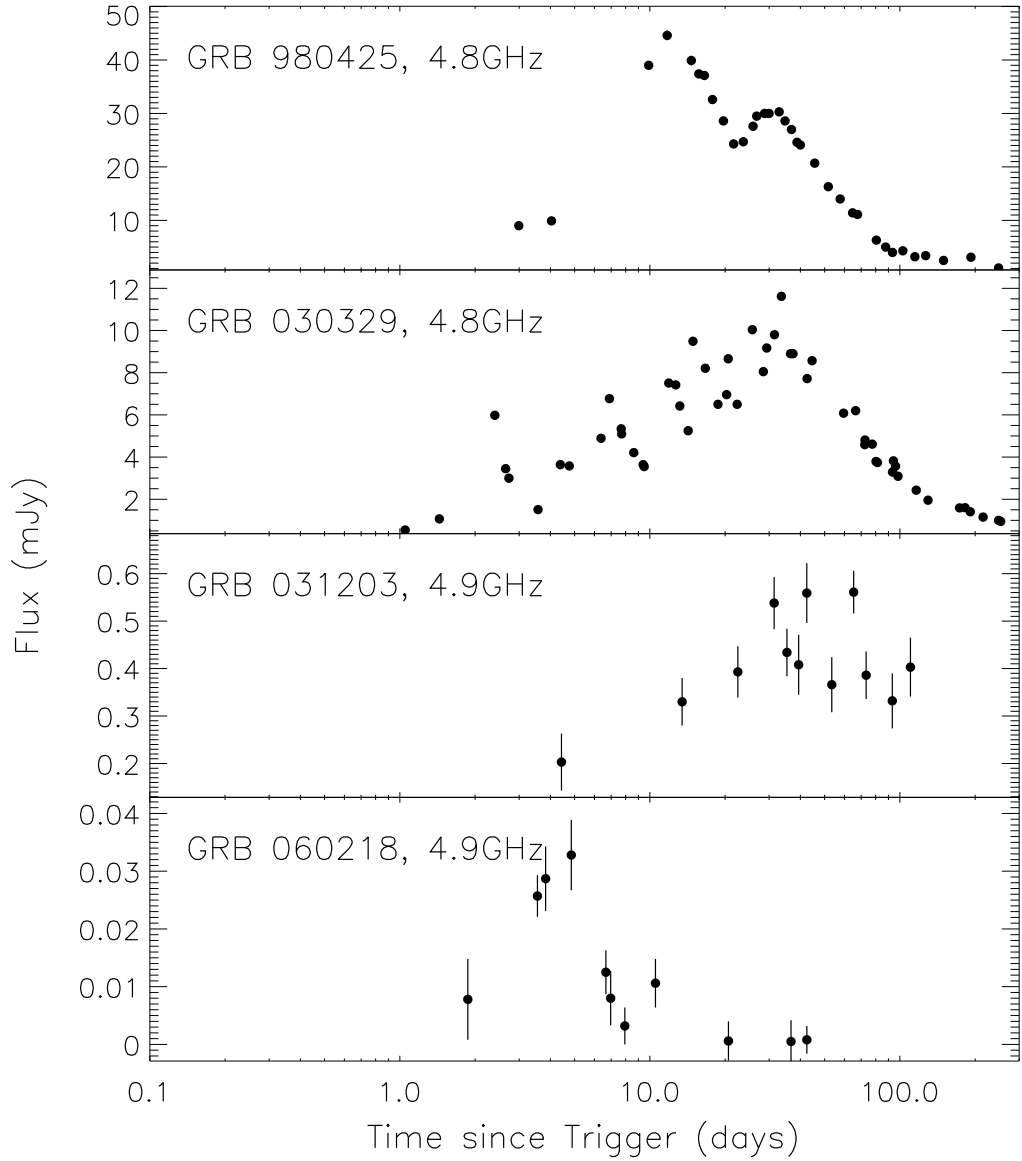}}
\caption{Radio lightcurves for all four events at 4.8/4.9 GHz. The errors 
associated with the data are very small for GRB\,980425 and GRB\,030329.}
\label{fig:radio_lc}
\end{figure}

We performed radio observations of GRB\,060218 with the Westerbork Synthesis 
Radio Telescope (WSRT); these results are tabulated in 
Table~\ref{table:radio-obs}. Additional radio data for this burst and radio 
data for the three other events were collected from the literature (GRB\,980425; 
\citealt{fra03} and \citealt{kul98}, GRB\,030329; \citealt{ber03}, 
\citealt{fra05}, \citealt{tay05}, \citealt{van05}, and \citealt{res05}, 
GRB\,031203; \citealt{sod04}, GRB\,060218; \citealt{sod06} and \citealt{kam06}). 
The radio flux lightcurves in 4.8/4.9\,GHz, at which all of the four events were
observed, are shown in Figure~\ref{fig:radio_lc}.

\begin{figure}
\centerline{
\plotone{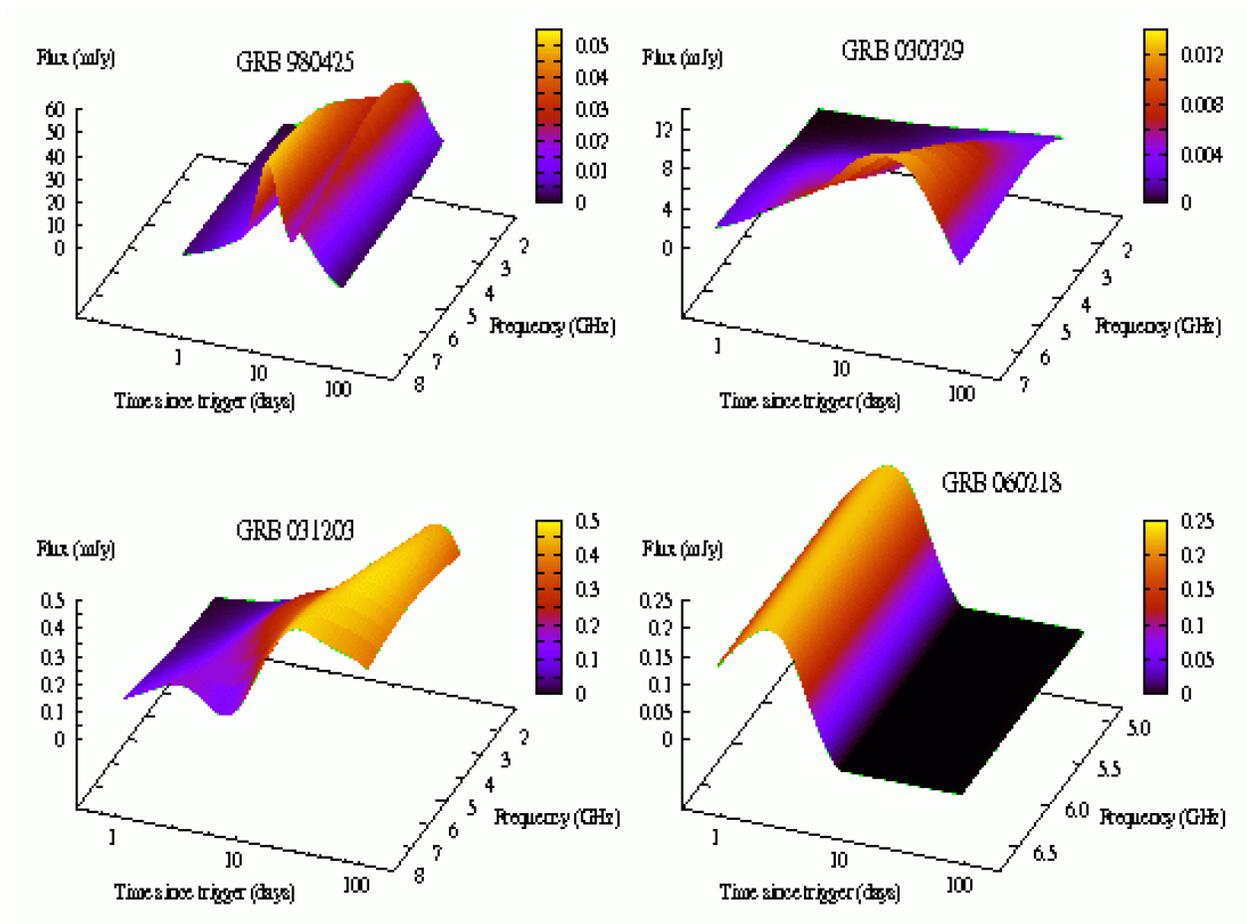}}
\caption{Spline fits to the broadband radio lightcurves for the four events.}
\label{fig:radio_3d}
\end{figure}

\begin{figure}
\centerline{
\plotone{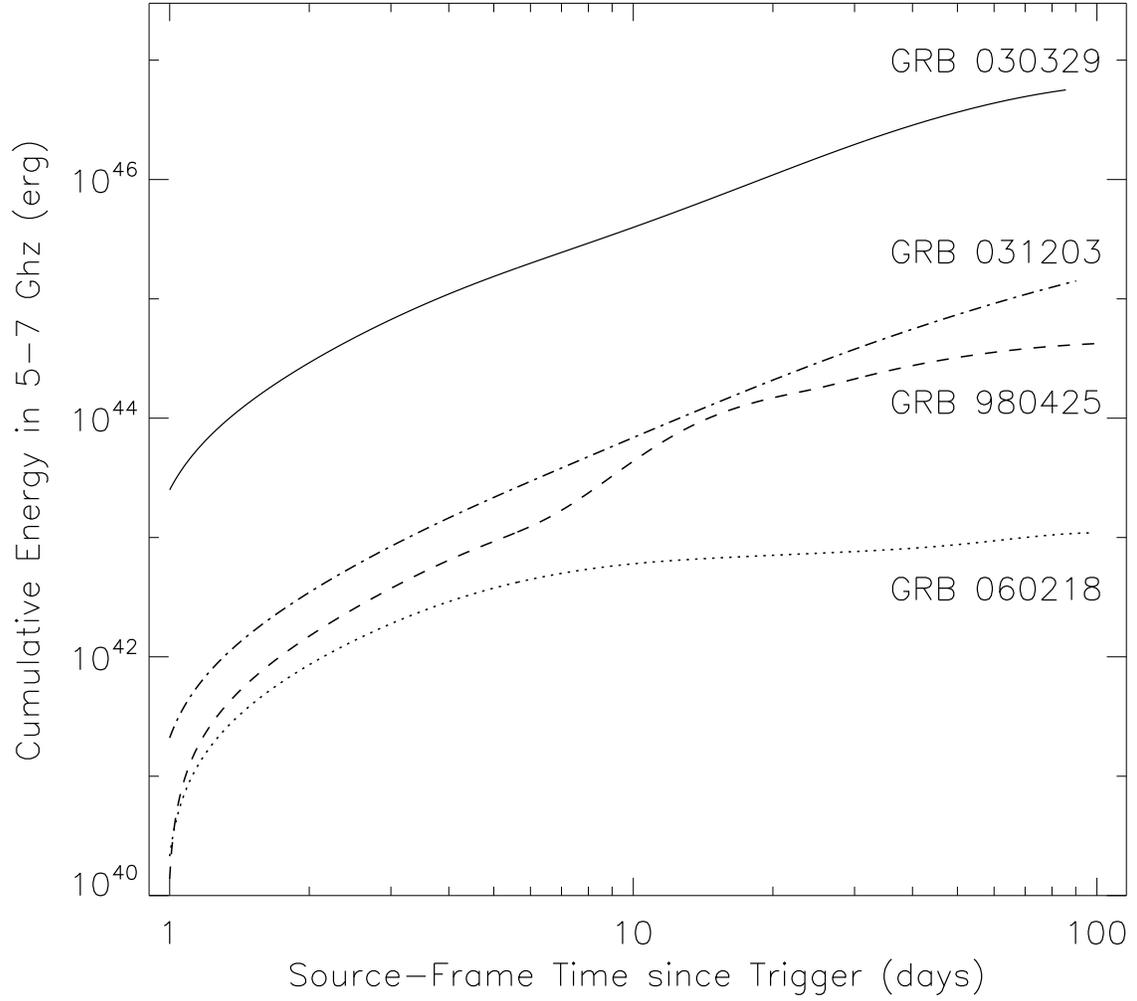}}
\caption{Cumulative isotropic-equivalent total emitted energy in 5$-$7 GHz 
(source-frame energy) as a function of source-frame time.}
\label{fig:lr_cum}
\end{figure}

As with the X-ray afterglow, we fitted the flux lightcurve of each event (in a 
wide band, where available) with a natural cubic spline, to estimate the 
cumulative energy evolution emitted in radio. To minimize the effect of 
scintillation in the radio lightcurves, we used fewer nodes than data points in 
the fitting. This resulted in a smooth fit to the data that retains the overall 
evolution of the lightcurve. The curve is forced to be 0 at very early times 
($<0.01\;$days) and late times ($>500\;$days), or to the inferred flux of the 
host galaxy in the case of GRB\,031203 \citep{sod04}. The fitted lightcurves were
then used to create a grid in time-frequency space to obtain the flux profile in
a wider frequency range (Figure~\ref{fig:radio_3d}). Using this, we constructed 
the cumulative radio afterglow energy evolution in 5$-$7\,GHz, shown in 
Figure~\ref{fig:lr_cum}. In Table~\ref{tab:comparison}, we also present the 
comparison of the radio afterglow properties of the four events, along with 
their $\gamma$-ray prompt and X-ray afterglow properties. Similar to $\xdur$ 
defined above, we define $\rdur$ here as the time in the source frame during 
which 90\% of the radiated afterglow energy is accumulated in 5$-$7\,GHz in the 
source-frame energy. The integration time of the radio afterglow emission is 
1\,day to 100\,days after the burst trigger time, in the rest frame of the 
source.

As can be seen in Table~\ref{tab:comparison}, the isotropic equivalent energy 
that is radiated in the radio ($\eriso$) is $\sim 3.5-5$ orders of magnitude 
smaller than that in X-rays, $\exiso$. This is predominantly due to the fact 
that $\nu F_\nu$ typically peaks closer to the X-rays than to the radio, and 
it is very flat above its peak while it falls much faster toward lower energies. 
Another effect that enhances the difference between $\eriso$ and $\exiso$ is 
that the latter is calculated over a much wider energy range ($0.3-10\;$keV 
{\it versus} $5-7\;$GHz). Finally, since these are isotropic equivalent 
energies, most of the contribution to $\eriso$ is from significantly later times
than for $\exiso$, and the collimation of the outflow together with relativistic 
beaming effects could result in much larger $\exiso$ than $\eriso$. We note that
for at least two (GRB\,908425 and GRB\,060218) of the SN-GRBs, the 
isotropic-equivalent emitted energy at optical wavelengths can be (much) larger 
than $\exiso$ because it is typically dominated by the contribution from the 
SNe (powered by radioactive decay) rather than by the GRB afterglow emission 
(see also \S~\ref{sec:SN}). 

\section{Prompt and Afterglow Properties}
\label{sec:comparison}

The isotropic-equivalent luminosity of GRB X-ray afterglows scaled to 
$t = 10$\,hr after the burst in the source frame, $\lxiso$(10\,hr), can be used 
as an approximate estimator for the energy in the afterglow shock for the 
following reasons. First, at $10\;$hr the X-ray band is typically above the two 
characteristic synchrotron frequencies, $\nu_m$ (of the accelerated electron 
with minimum energy) and $\nu_c$ (of the electron whose radiative cooling time 
equals the dynamical time), so that the flux has very weak 
dependence\footnote{When synchrotron-self Compton is taken into account the 
dependence on $\epsilon_B$ becomes much stronger \citep{GKP06}.} on $\epsilon_B$
and no dependence on the external density, both of which are associated with 
relatively large uncertainties. Second, at $10\;$hr the Lorentz factor of the 
afterglow shock is sufficiently small ($\Gamma\sim 10$) so that a large fraction
of the jet is visible (out to an angle of $\sim \Gamma^{-1} \sim 0.1\;$rad 
around the line of sight) and local inhomogeneities on small angular scales are 
averaged out. Finally, the fact that the ratio of $\lxiso$(10\,hr) and $\egiso$ 
is fairly constant for most GRBs, suggests that both can serve as a reasonable 
measure of the isotropic-equivalent energy content of the ejected outflow.

\begin{figure}
\centerline{
\plotone{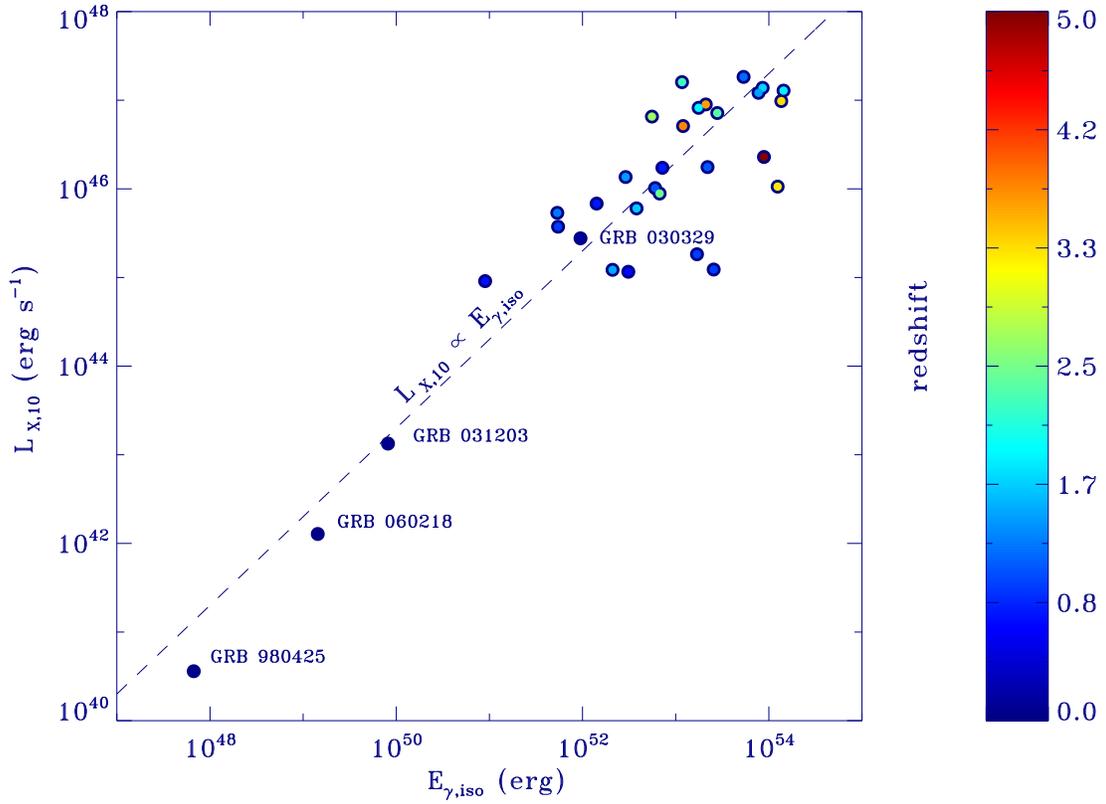}}
\caption{Isotropic-equivalent luminosity of SN-GRB X-ray afterglows
scaled to t = 10 hr (source frame; 2$-$10 keV) after the burst trigger as a 
function of their isotropic $\gamma$-ray energy release (20$-$2000 keV).
The redshift for each event is also shown in color
\citep[adopted from][]{nou06}.}
\label{fig:lx_eiso}
\end{figure}

Figure~\ref{fig:lx_eiso} shows $\lxiso$(10\,hr) in 2$-$10 keV rest-frame
energy as a function of their isotropic $\gamma$-ray energy release ($\egiso$,
in 20$-$2000 keV) for the four events, in comparison to regular GRBs. A 
linear relation, $\lxiso(10\,{\rm hr}) \propto \egiso$, seems to be broadly 
consistent with the data, probably suggesting a roughly universal efficiency 
for converting kinetic energy into gamma rays in the prompt emission for 
these four events and those of ``regular" long GRBs. This ``universal" 
efficiency is also likely to be high (i.e., the remaining kinetic energy is 
comparable to, or even smaller than, the energy dissipated and radiated in 
the prompt emission). If this is the case, the well-known efficiency problem 
for long GRBs also persists for SN-GRB events.
Not surprisingly, there is also a correlation between the cumulative X-ray 
afterglow energy and isotropic $\gamma$-ray energy (see 
Figure~\ref{fig:egxisot90} and also \S~\ref{sec:dis}).

\section{SN Properties}
\label{sec:SN}

\begin{figure}
\centerline{
\plotone{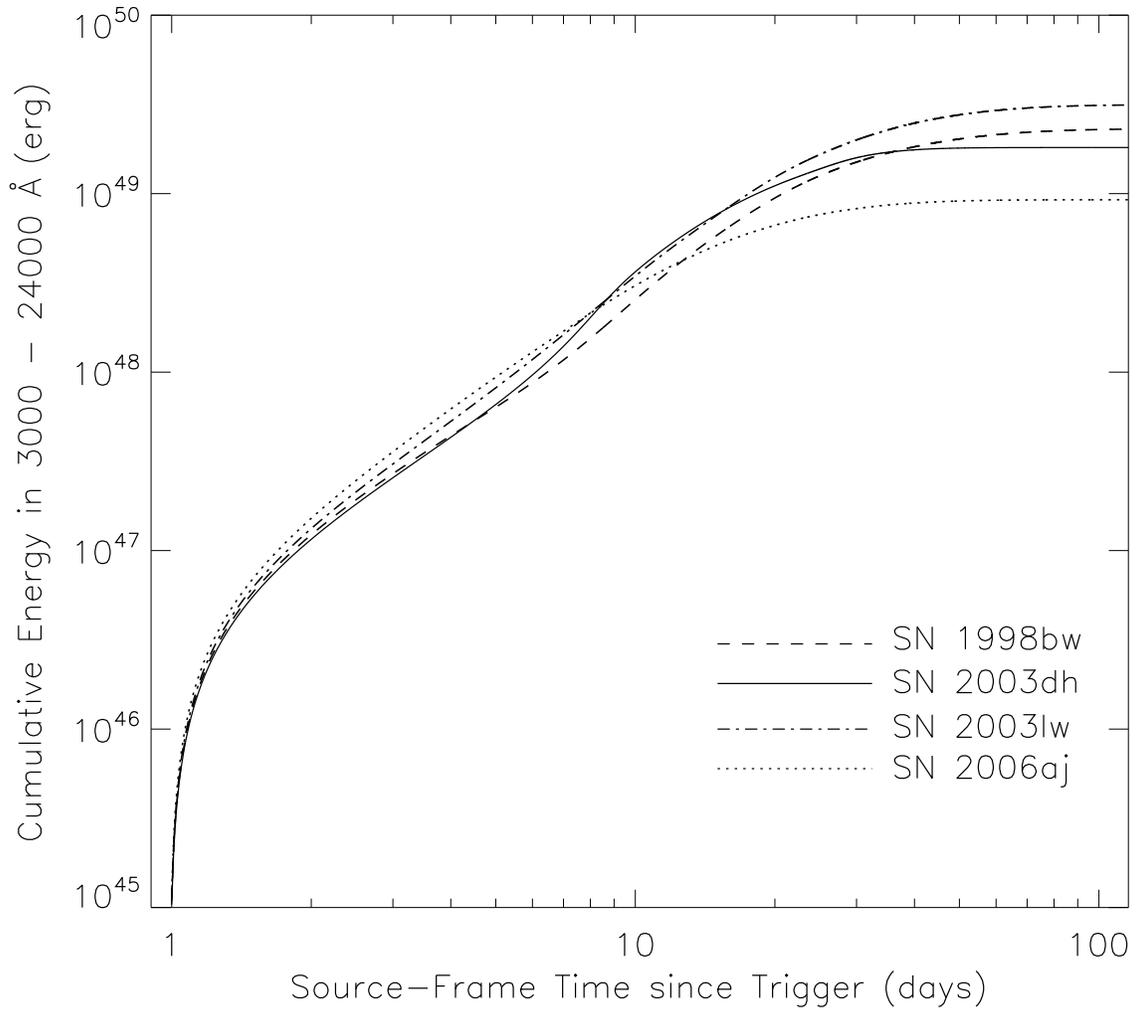}}
\caption{Cumulative isotropic-equivalent total emitted energy in 
3000$-$24000\,\AA~(source-frame energy) as a function of source-frame time, 
for all SNe associated with the GRBs presented in this work.}
\label{fig:sn_cum}
\end{figure}

\citet{pia06} presented the bolometric luminosity evolution (in optical and 
near-infrared wavelengths; 3000$-$24000\,\AA) for all the SNe associated with 
the GRBs considered here, namely SNe~1998bw, 2003dh, 2003lw, and 2006aj.  
We fitted a cubic spline to the luminosity evolution of each SN 
\citep[Figure 3 of][]{pia06}, and estimated the cumulative energy emitted by 
the SN as a function of source-frame time. To prevent the spline fits to
diverge after the last real data points ($\leq$ 40 days), we assumed that the
luminosity becomes much smaller ($\ll 10^{41}\;$erg~s$^{-1}$) at much later time 
($\gg 1000\;$days). The cumulative SN energy history is 
plotted in Figure \ref{fig:sn_cum} for all four SNe. We also estimated $\odur$,
during which 90\% of the SN energy is accumulated in 3000$-$24000\,\AA~since 1 
day after the burst, as well as the total energy emitted in the same energy 
range, $\eoiso$. These values are presented in Table~\ref{tab:comparison}, 
along with the burst properties in other wavelengths. It should be noted here 
that in most cases, $\eoiso$ is monopolized by the SN emission while the GRB
afterglow dominates at all other wavelengths. The integration time for the SN 
cumulative energy was from 1 day to 100\,days after the burst trigger. The 
cumulative energy of the four SNe span a much narrower range than the one 
reported for the entire SNe Ic class \citep{pam+06b}. Among these four, 
SN\,2006aj is the least energetic, albeit more energetic than other broad-lined 
SNe and normal SNe \citep{pia06}.

Furthermore, from the radio observations of type Ib/c SNe presented by 
\citet{wei02}, we note that the peak radio luminosity of SN\,1998bw observed at 
6\,cm (5\,GHz) was a few orders of magnitude higher than the other type Ib/c 
SNe, although the overall time evolution of this event was comparable to the 
others.

\section{Discussion}
\label{sec:dis}

\begin{figure}
\centerline{
\plotone{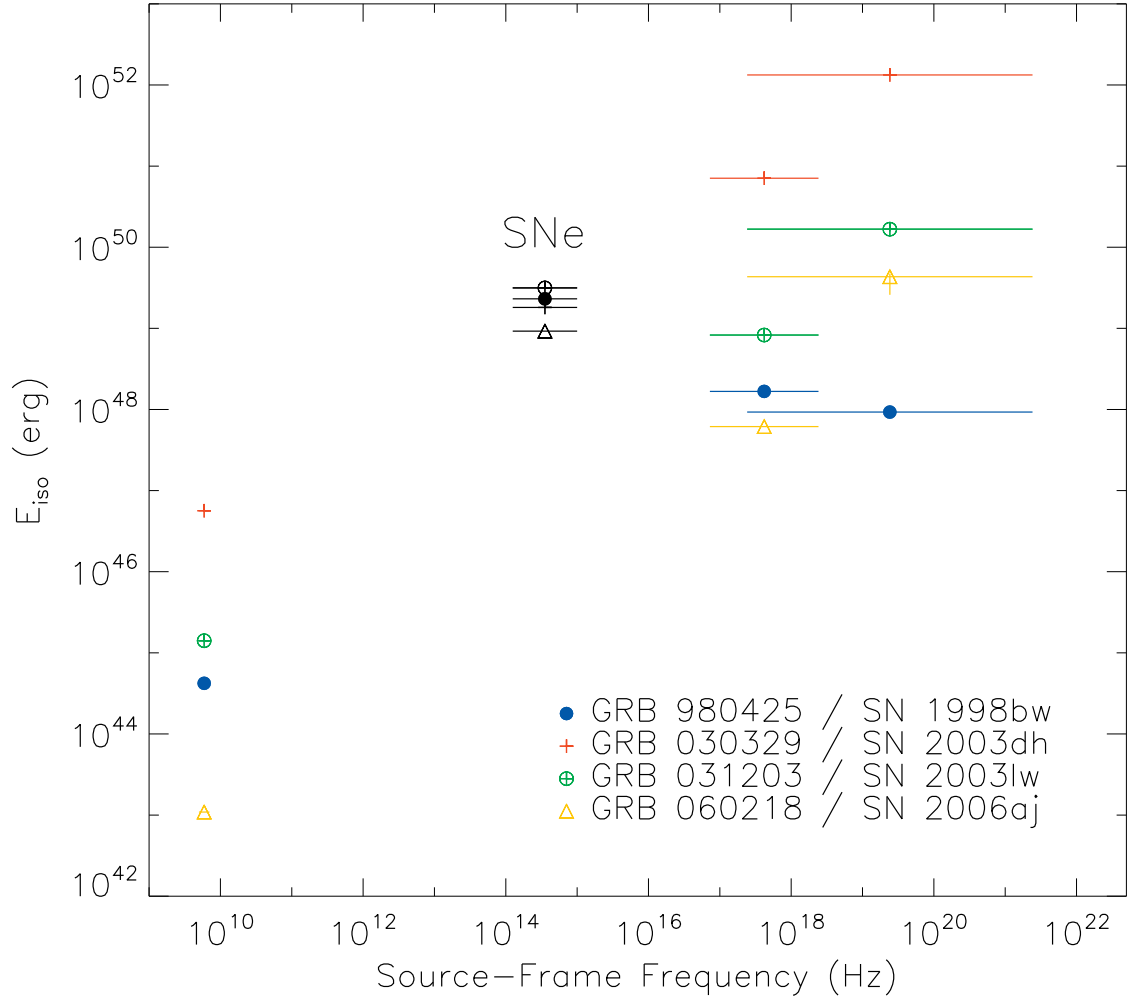}}
\caption{Summary of the isotropic-equivalent total emitted energy of the prompt 
and afterglow emission of the four GRBs, along with the properties of their 
associated SNe. The energetics in $\gamma$-ray, X-ray, Radio, and Optical (for
SNe) wavelengths correspond to 1$-$10,000\,keV, 0.3$-$10\,keV, 5$-$7\,GHz, 
3000$-$24000\,\AA~in the source frame, respectively.}
\label{fig:summary}
\end{figure}

The compilation of the radiated energy inventory, presented in 
Table~\ref{tab:comparison} and Figure~\ref{fig:summary}, offers an overview of 
the integrated effects of the energy transfers involved in all the physical 
processes of long GRB evolution, operating on scales ranging from AU to parsec 
lengths. The compilation also offers a way to assess how well we understand the 
physics of GRBs, by the degree of consistency among related entries. We present 
here our choices for the energy transfers responsible for the various entries in
the inventory. Many of the arguments in this exercise are updated versions of 
what is in the literature, but the present contribution is the considerable
range of consistency checks, which demonstrates that many of the entries in the 
inventory are meaningful and believable.

In the following, \S~\ref{gamma} repeats the minimal Lorentz factor estimates 
in \citet{LS01} using the parameters inferred for all of the four SN-GRB events 
in this work. \S~\ref{aftint} compares a simple model for the emitted radiation 
to the observational constraints on the integrated energy radiated in the 
afterglow phase. Finally, the energy contents in relativistic and 
sub-relativistic form are estimated in \S\S~\ref{relativistic} and 
\ref{subrelativistic}, respectively.

\subsection{Constraints on the Lorentz Factor}
\label{gamma}
As is well known, the requirement that GRBs are optically thin to high-energy 
photons yields a lower limit on the Lorentz factor of the expansion. We 
estimate here the minimal Lorentz factor of the outflow based on our 
observational analysis, by using the results of \citet{LS01}. They derived two 
limits; A and B. Limit A is from the requirement that the optical depth to pair 
production in the source for the photon with the highest observed energy, 
$\mathcal{E}_{\rm max}$, should be smaller than unity. Limit B is from the 
requirement that the Thompson optical depth of the $e^\pm$ pairs that are 
produced by the high-energy photons in the source does not exceed unity. The
required parameters to estimate these limits are the variability time of the 
$\gamma$-ray lightcurve, $\delta T$ (which is related to the radius of emission 
by $R \sim \Gamma^2 c\delta T$), and the photon flux per unit energy, 
$dN_{\rm ph}/dAdtd\mathcal{E}_{\rm ph} = f\mathcal{E}_{\rm ph}^{-\beta}$ for 
$\mathcal{E}_{\rm min} < \mathcal{E}_{\rm ph} < \mathcal{E}_{\rm max}$: in the 
latter, both the normalization $f$ and the high energy photon index $\beta$ are
needed. Although the result depends on the exact choice of parameters, 
representative values are presented in Table~\ref{Table:Gamma_min}. For each 
event, we determined $\delta T$ to be the Full Width Half Maximum (FWHM) of a 
two-sided Gaussian function fitted to the prompt $\gamma$-ray lightcurve, the 
peak of which aligns with the main peak of the lightcurve. We then analyzed the 
``peak'' spectrum with the integration time of $\delta T$ to obtain the 
high-energy photon index $\beta$ and the $\epeak$ values shown here. The minimal
values of the Lorentz factor derived in this way are generally smaller for less
energetic events (in terms of $\egiso$ or $\exiso$). For the event with the 
smallest $\egiso$, GRB\,980425, we find $\Gamma_{\rm min} \sim 2.4$ if we adopt 
our lower limit for the high-energy photon index ($\beta > 3.5$; 1$\sigma$), 
while a larger value of $\beta$ would lower $\Gamma_{\rm min}$ \citep[consistent
with the results of][]{LS01}.

\subsection{Integrated Radiated Energy}
\label{aftint}

The integrated (isotropic-equivalent) radiated energy during the afterglow, is 
given by
\begin{eqnarray}\nonumber
E_{\rm rad,iso} &=& \int dt \int d\nu \,L_{\rm\nu,iso}(t) \\ &=&
\eta_t t_{\rm peak}L_{\rm iso}(t_{\rm peak}) = \eta_t\eta_\nu t_{\rm
peak}\nu_{\rm peak}(t_{\rm peak})L_{\rm\nu,iso}(t_{\rm peak})\ ,
\end{eqnarray}
where $t$, $\nu$, and $L_{\rm\nu,iso}$ are measured in the cosmological frame 
of the GRB, $t_{\rm peak}$ is the time when $tL_{\rm iso}(t)$ peaks, $\nu$ is 
the frequency where $\nu L_{\rm\nu,iso}(t_{\rm peak})$ peaks;
\begin{equation}
\eta_t = {\int dt\,L_{\rm
iso}(t) \over t_{\rm peak}L_{\rm iso}(t_{\rm peak})}\ ,
\end{equation}
and 
\begin{equation}
\eta_\nu = {L_{\rm iso}(t_{\rm peak}) \over \nu_{\rm peak}(t_{\rm
peak})L_{\rm\nu,iso}(t_{\rm peak})}\ ,
\end{equation}
are factors of order unity here (although we typically expect 
$\eta_t\eta_\nu \sim 10$). Since in practice $\nu L_{\rm\nu,iso}$ peaks near the
X-ray band, we can assume that $\lxiso$ only mildly underestimates\footnote{Even
if $\nu F_\nu$ peaks below the X-rays, it is very flat above its peak, so a 
significant fraction of the afterglow energy is still radiated in the X-rays.} 
$L_{\rm iso}$, and the time $t_{\rm peak}$ when $tL_{\rm iso}(t)$ peaks is 
usually rather close to the time $t_{\rm X}$ when $t \lxiso(t)$ peaks. 
Therefore, $t_{\rm X} \lxiso(t_{\rm X})$ provides a convenient lower limit for 
$E_{\rm rad,iso}$ within an order of magnitude, although it is still possible 
that the X-ray observations might have missed the actual time when most of the 
energy was radiated, resulting in a significant underestimate. The values of 
$t_{\rm X} \lxiso(t_{\rm X})$ are provided in Table~\ref{Table:E_AG}. We find 
that these values are typically a factor of $\sim 2-3$ smaller than $\exiso$, 
suggesting $\eta_t \sim 2-3$. Although $\eta_t$ is defined for the bolometric 
luminosity rather than for the X-ray luminosity, the values derived here are 
still fairly representative.

\subsection{Energy Inventory - Relativistic Form}
\label{relativistic}

Supernova remnants are understood reasonably well, despite continuing 
uncertainty about the initiating explosion; likewise, we hope to understand the 
afterglow of GRBs, despite the uncertainties about their trigger mechanism. The 
simplest hypothesis is that the afterglow is due to a relativistic expanding 
blast wave.\footnote{For GRB\,030329 this picture is supported by direct 
measurements of the angular size of its radio afterglow image \citep{Taylor04,
tay05}, which show a superluminal apparent expansion velocity that decreases 
with time, in good agreement with the predictions of afterglow models 
\citep{ONP04,GR-RL05}.} The complex time structure of some bursts suggests that 
the central trigger may continue (i.e., the central engine may remain active) 
for up to $\sim 100$ seconds.  However, at much later times all memory of the 
initial time structure would be lost. All that matters then is essentially how 
much energy has been injected and its distribution in angle and in Lorentz 
factor, $\epsilon(>\Gamma,\theta)$, where $\epsilon \equiv dE/d\Omega$.

\subsubsection{Kinetic Energy Content}

An accurate estimate of the kinetic energy in the afterglow shock requires 
detailed afterglow modeling and good broadband monitoring, which enables one to 
determine the values of the shock microphysical parameters (electron and 
magnetic energy equipartition fractions, $\epsilon_e$ and $\epsilon_B$, and 
shock-accelerated electron power-law index, $p$). However, even then, it 
provides only a lower limit for the kinetic energy due to the conventional and 
highly uncertain assumption that all of the electrons are accelerated to
relativistic energies \citep{EW05,GKP06}. Nevertheless, an approximate lower 
limit on the isotropic-equivalent kinetic energy in the afterglow shock, 
$\ekiso$, can be obtained from the isotropic-equivalent X-ray luminosity, 
$\lxiso$, since the typical efficiency of the X-ray afterglow, $\epsilon_{\rm X}
\equiv t\lxiso(t)/\ekiso(t)$, is $\lesssim 10^{-2}$ \citep{GKP06}: a rough
lower limit on $\ekiso$ is obtained by adopting $\epsilon(t_{\rm X}) \sim 
10^{-2}$, where $t_{\rm X}$ is defined in the previous section and the values 
for the four events are presented in Table~\ref{Table:E_AG}. For GRBs~980425
and 060218, we estimate $\ekiso(t_{\rm X}) \approx 6\times 
10^{49}\epsilon_{\rm X,-2}^{-1}\;$erg, and $\ekiso(t_{\rm X}) \approx 3\times 
10^{49}\epsilon_{\rm X,-2}^{-1}\;$erg, respectively,
where $\epsilon_{\rm X,-2} = \epsilon_{\rm X}(t_{\rm X})/10^{-2}$. 
These rough estimates are similar to the energy inferred from a more detailed 
analysis of the X-ray and radio observations of this event 
\citep[$\ekiso\approx 5\times 10^{49}\;$erg;][]{Waxman04}. 
For GRB~031203, we find $\ekiso(t_{\rm X}) \approx 2\times 
10^{50}\epsilon_{\rm X,-2}^{-1}\;$erg.
    
For GRB\,030329, we derive $\ekiso(t_{\rm X}) \approx 5\times
10^{52}\epsilon_{\rm X,-2}^{-1}\;$erg. This estimate is comparable to that from 
the broadband spectrum at $t \approx 10\;$days 
\citep[$\ekiso\sim 5\times10^{52}\;$erg for a uniform density and 
$\sim 10^{52}\;$erg for a wind;][]{GR-RL05}, assuming negligible lateral
expansion of the jet \citep{goro}. For rapid lateral expansion, the inferred 
value of $\ekiso(10\;{\rm days})$ is lower 
\citep[$\sim 1.6\times10^{51}-5\times10^{51}\;$erg,][]{ber03,GR-RL05} but should 
correspond to a similar $\ekiso$ before the jet break time ($t_j \approx
0.5\;$days), for a comparable initial half-opening angle. There were, however, 
several re-brightening episodes observed in the optical afterglow lightcurve of 
GRB\,030329, between $\sim 1.5\;$days and a week after the GRB \citep{Lipkin04}, 
suggesting energy injection into the afterglow shock that increased its energy 
by a factor of $\sim 10$ \citep{GNP03}. This would imply that $\ekiso$ at 
$t_{\rm X} \approx 2\times 10^4\;$s was a factor of $\sim 10$ lower than our 
rough estimate, or that $\epsilon_{\rm X}(t_{\rm X})$ is as high as $\sim 0.1$. 
A possible alternative explanation for the relatively high value of $t_{\rm X} 
\lxiso(t_{\rm X})$ which does not require a high afterglow efficiency 
($\epsilon_{\rm X}$) comes about if a flare in the X-rays was present around 
$t_{\rm X}$, which was not detected in the optical lightcurve at the similar 
time. In this case, $\lxiso(t_{\rm X})$ would be dominated by late time 
activity of the central source rather than by emission from the external shock
\citep{RR01,GNP03}.

\subsubsection{Minimal Energy Estimates}

Here we derive a simple but rather robust estimate for the minimal combined 
energy in the magnetic field ($E_B$) and in the relativistic electrons ($E_e$) 
that are responsible for the observed synchrotron emission of flux density 
$F_{\nu,R}$ at some frequency $\nu_R$, $\emin = \min(E_B+E_e)$. This estimate 
is applied for the sub-relativistic flow, to avoid the effects of relativistic 
beaming and reduce the uncertainty on the geometry of the emitting region. 
Since the source is not resolved, the energy is estimated near the time of the
non-relativistic transition, $t_{\rm NR}$, where we have a handle on the source 
size.

We follow standard equipartition arguments \citep{P70,SR77,Gaensler05,NPS05}.  
The minimal energy is obtained close to equipartition, when $E_B/E_e = 3/4$. At 
such late times ($t \sim t_{\rm NR} \gtrsim 10^2\;$days) it is easiest to detect
the afterglow emission in the radio, so $\nu_R$ would typically be in the radio 
band. Furthermore, $\nu_m(t_{\rm NR})$ is also usually around the radio band, 
meaning that the electrons radiating in the radio would carry a reasonable 
fraction of the total energy in relativistic electrons. Still, the total energy 
of all electrons would be a factor of $\gtrsim 10$ larger than $E_e$, which 
would increase the total minimal energy by a factor of $\gtrsim 6$. Since the 
kinetic energy is expected to be at least comparable to that in the relativistic
electrons and in the magnetic field, the total energy is likely to be at least 
an order of magnitude larger than $\emin$.

Following \citet{NPS05}, $E_e = N_e\gamma_e m_e c^2$, where $N_e$ is the number 
of electrons with a synchrotron frequency $\nu_{\rm syn} \sim \nu_R$, and 
therefore, with a Lorentz factor $\gamma_e \approx [2\pi m_e c 
(1+z)\nu_R/eB]^{1/2}$. Also, since $F_{\nu,R} \approx N_e P_{\nu,{\rm max}}
(1+z)/4\pi d_L^2$ where $P_{\nu,{\rm max}} \approx P_{\rm syn}(\gamma_e)/
\nu_{\rm syn}(\gamma_e) \approx \sigma_T m_e c^2 B/3e$, $N_e \approx 12\pi 
d_L^2 e F_{\nu,R}/(1+z)\sigma_T m_e c^2 B$. Estimating the emitting volume as 
$V = (4\pi/3)R^3/\eta$ with $\eta = 10\eta_1$ ($\eta^{-1}$ is the fraction of 
the volume of a sphere with the radius $R$ of the emitting material that is actually occupied by the
emitting material), while in terms of our observed quantity, $t_{\rm NR}$, 
$R(t_{\rm NR}) = act_{\rm NR}/(1+z)$ with $a = 2a_{0.3}$ ($a$ is the average 
apparent expansion velocity at $t_{\rm NR}$ in the cosmological frame of 
the GRB/SN in units of the speed of light), we obtain
\begin{equation}\label{E_min}
\emin = 6\times 10^{49}a_{0.3}^{9/7}\eta_1^{-3/7}(1+z)^{-19/14}
d_{L28}^{8/7}\left(\frac{F_{\nu,R}}{1\;{\rm mJy}}\right)^{4/7}
\left(\frac{\nu_R}{5\;{\rm GHz}}\right)^{2/7} \left(\frac{t_{\rm
NR}}{100\;{\rm days}}\right)^{9/7}\ {\rm ergs}.
\end{equation}
The resulting estimates of $\emin$ for the four events discussed in this paper 
are given in Table~\ref{Table:E_min}, along with $t_{\rm NR}$, and $F_{\nu,R}$
values.  We used $\nu_R = 4.86$\,GHz.
As a sanity check, we calculate the minimal external density 
that corresponds to a total energy of $10\emin$; $n_{\rm min} = 10\emin/(4\pi/3)
R_{\rm NR}^3 m_p c^2$, where $R_{\rm NR} = R(t_{\rm NR}) = ac t_{\rm NR}/(1+z)$,
in which we use our fiducial values of $a = 2$ and $\eta = 10$. These values
are also presented in Table~\ref{Table:E_min}. It is useful to compare $\emin$ 
to other energy estimates. For GRB\,980425, the X-ray afterglow observations 
suggest an energy of $\sim 5\times 10^{49}\;$erg $\sim 30 \emin$ in a mildly 
relativistic roughly spherical component \citep{Waxman04}. For GRB\,030329 the 
total kinetic energy at late times is estimated to be $\sim 3.2\times 
10^{50}\;$erg $\sim 100 \emin$ \citep{GR-RL05}. Therefore, it can be seen that 
while the ratio of the total kinetic energy in relativistic outflow ($\gamma\beta 
> 1$) around $t_{\rm NR}$ and $\emin$ is indeed $\gtrsim 10$, it is typically 
$\lesssim 100$. Moreover, the fact that these different energy estimates are 
consistent lends some credence to these models.

Some cautionary remarks are in order. The above calculation is only sketchy and 
should be taken as an order of magnitude estimate at present. For example, the 
usual assumption that at $t_{\rm NR}$ the flow is already reasonably well 
described by the Newtonian spherical Sedov-Taylor self-similar solution, for 
which $\eta \approx 10$, is probably not a very good approximation. Numerical 
studies show that there is very little lateral expansion of the GRB jets while 
the flow is relativistic, and therefore it takes at least several dynamical
timescales after $t_{\rm NR}$ for the flow to approach spherical symmetry. 
Furthermore, since the flow is still mildly relativistic at this stage, there 
is still non-negligible relativistic beaming of the radiation toward the 
observer from the forward jet and away from the observer from the counter-jet. 
Altogether, the flux is somewhat enhanced due to this mild beaming, and the 
fraction of the total solid angle that is occupied by the flow at $t_{\rm NR}$ 
is still considerably smaller than unity. Consequently, this introduces an
uncertainty of at least a factor of a few in the estimate of $\emin$; however, 
since the total energy is expected to be $\gtrsim 10\emin$, it would still be at 
least a factor of a few larger than our estimate of $\emin$. The theoretical 
uncertainty on the dynamics of the flow should improve with time as more 
detailed numerical studies become available.

Another important uncertainty is in the determination of the non-relativistic 
transition time, $t_{\rm NR}$. The better estimate of $\emin$ should become
obtainable as more well-sampled afterglow observations are made, and the 
modeling gets more precise so that one can more carefully estimate both 
$t_{\rm NR}$ and $F_{\nu,R}(t_{\rm NR})$. For GRB\,980425 we related $t_{\rm NR}$
to the time at which the X-ray lightcurve steepens, which likely corresponds to 
the deceleration time of the mildly relativistic ejecta. For GRB\,030329, 
$t_{\rm NR}$ was selected to roughly correspond to the time at which the radio
lightcurve flattens and at the same time to be consistent with estimates derived
from direct size measurements of the event \citep{ONP04,GR-RL05}. In the other 
two cases, only a crude estimate of $t_{\rm NR}$ can be made as it is unclear 
whether current observations clearly display a signature of the non-relativistic
transition.

\subsection{Energy Inventory - Sub-relativistic Form}
\label{subrelativistic}

Despite the wide range in energies in relativistic ejecta, and even wider range 
in $\egiso$, the total (non-neutrino) energy of the associated SN in all four 
events spans, at most, a factor of 10. Most of this energy is in 
non-relativistic ($\gamma \beta < 2$) kinetic energy: the integrated light of 
the SN is negligible. While not standard candles, the optical luminosities of 
the four SNe at peak are all much brighter than average Type Ib or Ic SNe 
\citep{woo06,pia06,fer06}. Since the brightness of Type I supernovae at peak is given 
by the instantaneous rate of decay of $^{56}$Ni, the $^{56}$Ni masses are thus 
inferred to be in the range of $\sim (0.2 - 0.7)\Msun$.

To produce this much $^{56}$Ni in a Wolf-Rayet (WR) star requires a kinetic 
energy of at least $\sim 2 \times 10^{51}$\,erg, even in lower mass WR stars 
\citep{Ens88}. In higher mass stars, a still greater energy is required for the 
lightcurve to peak within two weeks after the maximum. Large kinetic energies 
are also inferred from detailed models of the explosion, especially the 
lightcurve and the velocity histories of spectral features \citep[see][and 
references therein]{woo06}. In summary, the supernova kinetic energies in the
four well-studied events almost certainly lie within the range $2 \times 
10^{51} - 2 \times 10^{52}$\,erg. In fact, the range of typical GRB-SNe may be 
much smaller, with the brightness at peak varying by no more than one magnitude 
in all four events and the kinetic energy in at least three of the four events 
(all but SN\,2006aj) within a factor of two of $1.5 \times 10^{52}$\,erg 
\citep{woo06}.

\begin{figure}
\epsscale{0.9}
\centerline{
\plotone{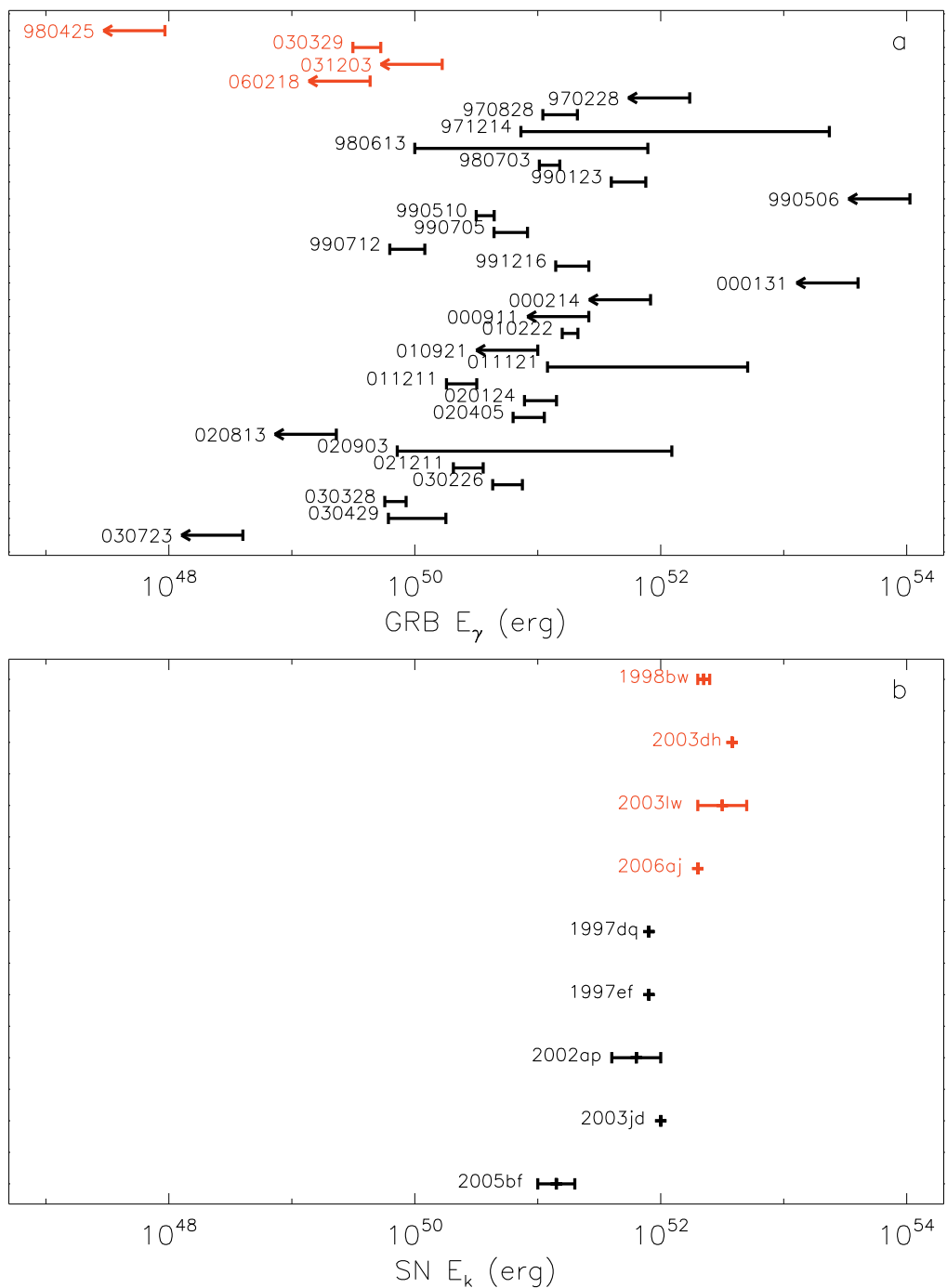}}
\caption{Comparisons of collimation-corrected total emitted $\gamma$-ray energy
($E_{\gamma}$) of the four SN-GRBs (red) and other GRBs (a), and of the SN kinetic 
energy ($E_{\rm k}$) of the four GRB-SNe (red) and other SNe of the same type (b).
In (a), $\egiso$ was used as an upper limit for GRBs with no jet angle
constraints.}
\label{fig:energetics}
\end{figure}

Finally, in Figure~\ref{fig:energetics}, we compare the collimation-corrected
total emitted energy in $\gamma$-ray ($E_\gamma$) and supernova kinetic energy
estimates ($E_{\rm k}$) of these four SN-GRB events with regular GRBs and other 
broad-lined (1998-bw like) SNe without GRBs.
For GRB\,030329, we used the jet angle estimate of $\theta_j \sim 0.083 - 
0.14\;$rad \citep{goro}. For the other three events, the isotropic-equivalent emitted
energy, $\egiso$, was used as an upper limit, as there was no observational
evidence of jet breaks for these events.
The $E_\gamma$ values for 27 regular GRBs were adopted from \citet{ghi04}, where
the $\egiso$ was again used as an upper limit for GRBs with no $\theta_j$
constraints.
All the $E_{\rm k}$ estimates were taken from the literature (SN\,1998bw;
\citealt{woo99}, SN\,2003dh; \citealt{maz03}, SN\,2003lw; \citealt{den05},
SN\,2006aj; \citealt{maz06}, SN\,2002ap; \citealt{maz02}, SN\,2003jd; 
\citealt{maz05}, SN\,1997dq \& SN\,1997dq; \citealt{maz04}, and SN\,2005bf; 
\citealt{tom05,fol06}).
We note that the explosion energy (i.e., SN $E_{\rm k}$) is much larger than 
the energy released as GRBs ($E_\gamma$), and spans much narrower range than
$E_\gamma$.

\section{Concluding Remarks}
\label{sec:summary}
One of the liveliest debated issues associated with GRBs is on the total energy 
released during the burster explosion: are GRBs standard candles? The GRB 
community has vacillated between initial claims that the GRB intrinsic 
luminosity distribution was very narrow \citep{h94}, to discounting all 
standard candle claims, to accepting a standard total GRB energy of $\sim 
10^{51}$ ergs \citep{f01}, and to diversifying GRBs into ``normal'' and 
``sub-energetic'' classes. The important new development is that we now have 
significant observational support for the existence of a sub-energetic 
population based on the different amounts of relativistic energy released during
the initial explosion. A network of theoretical tests lends credence to this
idea. The existence of a wide range of intrinsic energies that we presented in
this work may pose challenges to using GRBs as standard candles -- it is 
also worth stating explicitly that, when viewed together, these four events fall
away from the Amati relation.

Our results are consistent with the emerging hypothesis that GRBs and XRFs 
share a common origin in massive WR stars. The central engine gives rise to a 
polar outflow with two components \citep{woo06}. One large angle outflow (the 
SN), containing most of the energy and mass, is responsible for exploding the 
star and producing the $^{56}$Ni to make the SN bright. Only a tiny fraction of 
the material in this component reaches mildly relativistic velocities, which is 
more narrowly focused. A second outflow component (the GRB jet) occupies a 
narrower solid angle, probably contains smaller energy (which can range from
comparable to much smaller), and most of its energy is in material with 
relativistic velocities (where the typical Lorentz factor of the material that 
carries most of the energy in this component can vary significantly between 
different SN-GRBs). After it exits the star, internal shocks within this jet and 
external shocks with the residual wind material around the star make the GRB or 
XRF and its afterglow. Apparently, the properties of the broad component are not
nearly so diverse as those of the core jet \citep{R-RM04,sod06,woo06}.

We have argued, using well-known arguments connected with parameters such as 
opacity and variability timescales, that these less-energetic events do not require a 
highly-relativistic outflow. Our best estimates of Lorentz factors, $\Gamma$, for these
events are in the range of 2$-$10. 
Indeed, it is much more difficult to produce a jet with very 
high Lorentz factor -- i.e., a high energy loading per baryon -- than with 
low Lorentz factor. 
A jet with low Lorentz factor could result even if a jet of relatively pure 
energy is produced, since it may be loaded with excess baryons by instabilities 
at its walls as it passes through the star, or if it does not precisely 
maintain its orientation \citep{R-RCR02,aloy,ZWH04}. The above suggest that GRBs 
made by jets with lower Lorentz factor should be quite common in the universe
\citep{GR04}.

Continued advances in the observations will surely yield unexpected revisions 
and additions in our understanding of GRBs in connection with SNe: currently,
we are attempting to draw large conclusions from limited observations of 
exceedingly complex phenomena. However, the big surprise at the moment is that
these SN-GRB events appear to be intrinsically different from and much more
frequent \citep{GR04,guetta,pia06} than luminous GRBs, which have been
observed in large numbers out to higher redshifts.

We are very grateful to Scott Barthelmy and Takanori Sakamoto for
their help with the {\it Swift} BAT data analysis, to Rob Preece and Michael
Briggs for their help with the BATSE-WFC joint analysis,
and to Ersin G\"o\u{g}\"u\c{s}
for helpful discussions.  We also thank the
WSRT staff, in particular Tony Foley.  This work is supported by IAS
and NASA under contracts G05-6056Z (YK) and through a
Chandra Postdoctoral Fellowship award PF3-40028 (ERR), by the
Department of Energy under contract DE-AC03-76SF00515 (JG), and by
PPARC (ER). SEW acknowledges support from NASA (NNG05GG08G), and the DOE 
Program for Scientific Discovery through Advanced Computing (SciDAC;
DE-FC02-01ER41176). 
RAMJW is supported by the Netherlands Foundation for Scientific Research (NWO) 
through grant 639.043.302. This paper benefited from collaboration through 
an EU-funded RTN, grant number HPRN-CT-2002-00294. The Westerbork Synthesis 
Radio Telescope is operated by ASTRON (Netherlands Foundation for Research in 
Astronomy) with support from NWO. 


\clearpage

\clearpage

\clearpage

\clearpage
\begin{table}
\caption{Summary of the spectral fit results of GRB\,980425. The data were 
fitted with a Comptonization model; all uncertainties are 1$\sigma$.
\label{tab:980425_results}}
\begin{center}
\begin{tabular}{ccccc}
\hline 
\hline \\[-3ex]
Time & $A$\tablenotemark{a} & $\epeak$ & $\alpha$ & $\chi^2/{\rm dof}$ \\[-2pt]
Interval &  & (keV) &   &  \\[2pt]
\hline \\[-2ex] 
A & 17.4 \e 3.6 & 175 \e 13 & --0.12 \e 0.22 & 134.4/141 \\ 
B & 10.7 \e 1.1 & 133 \e ~8 & --1.16 \e 0.09 & 149.9/141 \\ 
C & ~4.2 \e 0.9 & ~34 \e ~3 & --1.54 \e 0.09 & 158.8/141 \\ 
D & ~2.0 $^{+~5.9}_{-~1.3}$ & ~14 $^{+~3}_{-~6}$ & --1.51 \e 0.36 & 121.8/141 \\ 
\hline 
\end{tabular}
\begin{minipage}[c]{5.5in}
\vspace{1ex}
\begin{footnotesize}
$^{\rm a}$~in units of $10^3$\,ph~s$^{-1}$\,cm$^{-2}$\,keV$^{-1}$.
\end{footnotesize}
\end{minipage}
\end{center}
\end{table}

\begin{table}
\caption{Summary of the spectral fit results of GRB\,031203 using a single 
power-law model; all uncertainties are 1$\sigma$.}
\label{tab:031203_results}
\begin{center}
\begin{tabular}{ccccc}
\hline 
\hline \\[-3ex]
Time & Amplitude\tablenotemark{a} & Photon & $\chi^2/{\rm dof}$ \\[-2pt]
Interval &  & Index &  \\[2pt]
\hline \\[-2ex] 
0 - 2 s  & ~6.56 \e 2.46 & --1.60 \e 0.09 & 11.2/7 \\ 
2 - 4 s  & 10.25 \e 4.26 & --1.75 \e 0.11 & ~4.9/7 \\ 
4 - 8 s  & ~4.81 \e 1.66 & --1.67 \e 0.09 & ~6.2/7 \\ 
8 - 50 s & ~0.72 \e 0.56 & --1.63 \e 0.20 & ~5.0/7 \\ 
\hline 
\end{tabular}
\begin{minipage}[c]{5.5in}
\vspace{1ex}
\begin{footnotesize}
$^{\rm a}$~in units of ph~s$^{-1}$\,cm$^{-2}$\,keV$^{-1}$, at 1\,keV.
\end{footnotesize}
\end{minipage}
\end{center}
\end{table}

\begin{table}
\begin{footnotesize}
\caption{Summary of the joint BAT-XRT time-resolved spectral fits for 
GRB\,060218. The spectra were fitted to a power law with high-energy cutoff
with a blackbody unless otherwise noted. Note that we present $\epeak$ instead 
of a cutoff energy. The uncertainties are 1$\sigma$.}
\label{tab:060218_results}
\begin{center}
\begin{tabular}{ccccccc}
\hline 
\hline \\[-3ex]
Time & $N_{\rm H}$ & $k$T & $\epeak$ & Photon
         & $\chi^2/{\rm dof}$ & Flux (0.5-150\,keV)\tablenotemark{a} \\[-2pt]
Interval & ($10^{22}\;$cm$^{-2}$) & (keV) &  (keV) & Index & & (10$^{-9}$\,erg\,s$^{-1}$cm$^{-2}$) \\[2pt]
\hline \\[-2ex] 
 --8 -- 140  s\tablenotemark{b} 
               &      ---     &      ---     & 24.9 \e 6.0 & --0.87 \e 0.75 &  52.4/57  & ~6.38 \e 4.00 \\ 
 140 -- 303  s & 0.47 \e 0.08 & 0.22 \e 0.05 & 36.1 \e 7.2 & --1.39 \e 0.08 & 238.3/292 & ~8.95 $^{+0.43}_{-1.40}$ \\ 
 304 -- 364  s & 0.66 \e 0.14 & 0.17 \e 0.03 & 20.8 \e 4.1 & --1.49 \e 0.11 & 165.7/190 & 12.84 $^{+0.49}_{-2.87}$ \\ 
 406 -- 496  s & 0.62 \e 0.11 & 0.17 \e 0.02 & 31.5 \e 5.0 & --1.39 \e 0.08 & 236.4/302 & 14.93 $^{+0.47}_{-1.85}$ \\ 
 496 -- 616  s & 0.58 \e 0.02 & 0.16 \e 0.02 & 22.3 \e 2.9 & --1.33 \e 0.06 & 323.6/381 & 13.99 $^{+0.47}_{-1.12}$ \\ 
 616 -- 736  s & 0.51 \e 0.07 & 0.20 \e 0.02 & 18.9 \e 2.6 & --1.21 \e 0.07 & 395.6/401 & 12.59 $^{+0.39}_{-1.02}$ \\ 
 736 -- 856  s & 0.64 \e 0.08 & 0.16 \e 0.02 & 15.2 \e 2.4 & --1.41 \e 0.07 & 397.1/412 & 13.40 $^{+0.37}_{-1.04}$ \\ 
 856 -- 976  s & 0.66 \e 0.08 & 0.15 \e 0.01 & 12.0 \e 2.0 & --1.45 \e 0.07 & 390.2/420 & 10.03 $^{+0.39}_{-1.37}$ \\ 
 976 -- 1256 s & 0.61 \e 0.06 & 0.15 \e 0.01 & ~5.7 \e 0.4 & --1.39 \e 0.09 & 611.7/567 & 11.17 $^{+0.23}_{-0.93}$ \\ 
1256 -- 1556 s & 0.65 \e 0.06 & 0.14 \e 0.01 & ~3.6 \e 0.2 & --1.30 \e 0.14 & 571.2/539 & 10.26 $^{+0.30}_{-0.97}$ \\ 
1557 -- 1857 s & 0.67 \e 0.06 & 0.13 \e 0.01 & ~2.4 \e 0.4 & --1.47 \e 0.03 & 483.5/481 & ~8.61 $^{+0.39}_{-0.95}$ \\ 
1857 -- 2157 s & 0.60 Fixed   & 0.14 \e 0.003 & ~1.8 \e 0.7 & --1.62 \e 0.15 & 427.8/433 & ~6.41 $^{+0.49}_{-0.62}$ \\ 
2157 -- 2457 s \tablenotemark{c}
               & 0.60 Fixed   & 0.14 \e 0.004 &    ---     & --2.45 \e 0.03 & 394.8/388 & ~6.34 $^{+0.42}_{-0.27}$ \\ 
2457 -- 2734 s \tablenotemark{c}
               & 0.60 Fixed   & 0.13 \e 0.004 &    ---     & --2.54 \e 0.04 & 327.1/342 & ~6.15 $^{+0.37}_{-0.52}$ \\ 
\hline 
\end{tabular}
\begin{minipage}[c]{5.5in}
\vspace{1ex}
\begin{footnotesize}
$^{\rm a}$~Unabsorbed flux. Uncertainties are associated with the absorbed flux 
            estimated from the fitted parameters. \\
$^{\rm b}$~Only BAT Event data were used. \\
$^{\rm c}$~Fitted by a power law.
\end{footnotesize}
\end{minipage}
\end{center}
\end{footnotesize}
\end{table}

\begin{table}
\caption{Summary of broadband properties of the prompt and afterglow emission 
of the four GRBs, along with the properties of their associated SNe. The 
energetics in $\gamma$-ray, X-ray, Radio, and Optical (for SNe) wavelengths 
correspond to 1$-$10,000\,keV, 0.3$-$10\,keV, 5$-$7\,GHz, 3000$-$24000\,\AA~in 
the source frame, respectively, unless noted.
\label{tab:comparison}}
\begin{center}
\begin{tabular}{ccccc}
\hline 
\hline \\[-2ex]
& 980425 & 030329 & 031203 & 060218 \\
& (1998bw) & (2003dh) & (2003lw) & (2006aj) \\[1ex]
\hline \\[-2ex]
$z$ & 0.0085 & 0.1685 & 0.105 & 0.0335 \\[1ex]
$\gdur$ (s) & 34.9 \e 3.8\tablenotemark{b} & 22.9\tablenotemark{c} 
   & 37.0 \e 1.3  & 2100 \e 100\tablenotemark{d} \\
$S_{\rm{X},2-30\;{\rm keV}}$\tablenotemark{a}  & 1.99E--6 & 6.71E--5 & 8.46E--7 & 1.09E--5 \\ 
$S_{\gamma,30-400\;{\rm keV}}$\tablenotemark{a} & 3.41E--6 & 1.20E--4 & 1.74E--6 & 3.09E--6 \\
$S_{\rm{X}}/S_{\gamma}$ & 0.58 & 0.56 & 0.49 & 3.54 \\
$\egiso$ (erg) & 9.29 \e 0.35 $\times 10^{47}$  
   & 1.33 $\times 10^{52}$ & 1.67 $^{+0.04}_{-0.10}$ $\times 10^{50}$ 
   & 4.33 $^{+0.41}_{-1.74}$ $\times 10^{49}$ \\
$\epeak$ (keV) & 122 \e 17 & 70 \e 2\tablenotemark{c} 
   & $> 71$ & 4.7 \e 1.2 \\[1ex] \hline \\[-2ex]
$\xdur$ (day) & 640.1 & 2.61 & 79.8 & 9.31 \\
$\exiso$ (erg) & 1.67 $\times 10^{48}$ & 7.09 $\times 10^{50}$ 
   & 8.27 $\times 10^{48}$ & 6.15 $\times 10^{47}$ \\[1ex] \hline \\[-2ex]
$\rdur$ (day) & 68.6 & 72.4 & 81.2 & 67.2 \\
$\eriso$ (erg) & 4.21 $\times 10^{44}$ & 5.64 $\times 10^{46}$ 
   & 1.41 $\times 10^{45}$ & 1.09 $\times 10^{43}$ \\[1ex] \hline \\[-2ex]
$\odur$ (day) & 53.9 & 31.4 & 53.6 & 30.7 \\
$\eoiso$ (erg) & 2.31 $\times 10^{49}$ & 1.81 $\times 10^{49}$
   & 3.15 $\times 10^{49}$ & 9.24 $\times 10^{48}$ \\[1ex]
\hline
\end{tabular}
\begin{minipage}[c]{5.5in}
\vspace{1ex}
\begin{footnotesize}
$^{\rm a}$~in units of erg~cm$^{-2}$. \\
$^{\rm b}$~Reference: BATSE current catalog 
   ({\url http://gammaray.nsstc.nasa.gov/batse/grb/catalog/current/}). \\
$^{\rm c}$~Reference: \citet{van04}. \\
$^{\rm d}$~Reference: \citet{cam06}.
\end{footnotesize}
\end{minipage}
\end{center}
\end{table}

\begin{table}
\tabcolsep = 4pt
\caption{\label{table:radio-obs}Log of the radio observations for GRB\,060218. 
The flux calibration has been performed using the source 3C286 as standard.
The observations on February 21 and February 28 were alternated between 1.4 
and 4.9\,GHz, and 2.3 and 4.9\,GHz, respectively, observing in 40 minute 
blocks at each frequency.}
\begin{center}
\begin{tabular}{lcccc@{ $\pm$ }l}
\hline 
\hline \\[-3ex]
\multicolumn{1}{c}{Date (2006)} &
\multicolumn{1}{c}{$\Delta$T} &
\multicolumn{1}{c}{Integration\,Time} &
\multicolumn{1}{c}{Frequency} &
\multicolumn{2}{c}{Flux} \\
\multicolumn{1}{c}{} &
\multicolumn{1}{c}{(days since trigger)} &
\multicolumn{1}{c}{(Hours)} &
\multicolumn{1}{c}{(GHz)} &
\multicolumn{2}{c}{($\mu$Jy)} \\
\hline \\[-2ex] 
Feb 21.452 - 21.951  &   3.30 -  3.80  &   5.9  &  1.4  &   $89$  &  $108$  \\
Feb 28.463 - 28.908  &  10.31 - 10.76  &   5.2  &  2.3  &  $-35$  &  $70$   \\
Feb 21.481 - 21.927  &   3.33 -  3.78  &   5.2  &  4.9  &  $257$  &  $36$   \\
Feb 24.688 - 24.943  &   6.54 -  6.79  &   5.9  &  4.9  &  $125$  &  $38$   \\
Feb 28.433 - 28.932  &  10.28 - 10.78  &   5.8  &  4.9  &  $106$  &  $42$   \\
Mar 10.642 - 10.905  &  20.49 - 20.76  &   6.1  &  4.9  &    $6$  &  $34$   \\
Apr  1.345 -  1.844  &  42.20 - 42.70  &  12.0  &  4.9  &    $8$  &  $24$   \\
\hline 
\end{tabular}
\end{center}
\end{table}

\begin{table}
\caption{Lower limits on the Lorentz factor of the material responsible 
for the prompt emission \citep[following the formalism of][]{LS01}. 
The high-energy photon index ($\beta$) and $\epeak$ values are for the peak 
spectra of duration $\delta T$, where the parameter limits are 1$\sigma$.}
\label{Table:Gamma_min}
\begin{center}
\begin{tabular}{cccccccccc}
\hline 
\hline \\[-3ex]
GRB &  & $\delta T$ &
$F_p$\tablenotemark{a} &  & $\epeak$
& $\mathcal{E}_{\rm max}$\tablenotemark{b} &  & Limit & 
Limit \\
 & \rb{$z$} & (s) & 
 & 
\rb{$\beta$} & (keV) & (keV) &
\rb{$\hat{\tau}$\tablenotemark{c}} & A & B \\
\hline \\[-2ex] 
980425 & 0.0085 & 12.6 & 0.00314          & $>$3.5 & 143 & 300 & 330 & 1.6 &
{\bf 2.4}  \vspace{0.2cm}\\
030329 & 0.1685 & 5.27 & 0.70              & 2.44 & 52.5 & 400 & $4.9\times
10^7$ & 13 & {\bf 27} \vspace{0.2cm}\\ 
 & & & & 2 & & & $1.0\times 10^6$ & 8.9 & {\bf 16} \\ 
\rb{031203} & \rb{0.105}  & \rb{8.16} & \rb{$<$0.002} & 3 &
\rb{$>$148} & \rb{200} &  $1.5\times 10^5$ & 3.7 & {\bf 7.5} \vspace{0.2cm}\\ 
060218 & 0.0335 & 1981 & 0.020 & 2.75 & 8.8  & 100 & 0.36 & --- & --- \\
\hline 
\end{tabular}
\begin{minipage}[c]{5.5in}
\vspace{1ex}
\begin{footnotesize}
$^{\rm a}$~The flux normalization at $\epeak$, $F_p = f\epeak^{-\beta}$,
in units of ph~s$^{-1}$\,cm$^{-2}$\,keV$^{-1}$. \\
$^{\rm b}$~The maximum photon energy with significant detection. \\
$^{\rm c}$~The optical depth for a photon of energy $m_e c^2$ for 
$\gamma\beta \approx 1$.
\end{footnotesize}
\end{minipage}
\tablecomments{For $\hat{\tau} < 1$ and
$\mathcal{E}_{\rm max} < m_e c^2$ (as is the case for GRB\,060218)
there is no lower limit on the Lorentz factor $\gamma$. Limit B derivation 
assumes that the high-energy power law continues significantly above 
the observed range for the cases presented in the Table. 
For GRB\,031203 there is only a lower limit on $\epeak$, and $\beta$ is not
measured, so we conservatively take $\epeak$ to be equal to its lower limit 
and show results for two representative values of $\beta$.}
\end{center}
\end{table}

\begin{table}
\caption{The source-frame time, $t_{\rm X}$, of the peak in the $t\lxiso$ 
history and the corresponding $t_{\rm X} \lxiso(t_{\rm X})$ values for the four
events.}
\label{Table:E_AG}
\begin{center}
\begin{tabular}{ccccl}
\hline 
\hline \\[-3ex]
GRB &  & $t_{\rm X}$ &
$t_{\rm X}\lxiso(t_{\rm X})$ & $t_{\rm X}\lxiso(t_{\rm X})$ \\
 & \rb{$z$} & (s) & (${\rm erg}$) & $/E_{\rm\gamma,iso}$ \\
\hline \\[-2ex] 
980425 & 0.0085 & $2.4 \times 10^7$ & $6.3 \times 10^{47}$ & 0.68  \\
030329 & 0.1685 & $1.7 \times 10^4$ & $5.3 \times 10^{50}$ & 0.040 \\
031203 & 0.105  & $3.5 \times 10^5$ & $2.1 \times 10^{48}$ & 0.013 \\
060218 & 0.0335 & $3.4 \times 10^4$ & $2.9 \times 10^{47}$ & 0.0068 \\
\hline 
\end{tabular}
\end{center}
\end{table}

\begin{table}
\caption{Summary of the estimated minimal combined energy, $\emin$, 
for all four events, with $\nu_R = 4.86$\,GHz.}
\label{Table:E_min}
\begin{center}
\begin{tabular}{cccccl}
\hline 
\hline \\[-3ex]
GRB & & $t_{\rm NR}$ & $F_{\nu,R}$ & $\emin$ & $n_{\rm min}$ \\
 & \rb{$z$} & (days) & (mJy) & (erg) & (cm$^{-3}$) \\
\hline \\[-2ex] 
980425 & 0.0085 & 200 & 3    & $1.5\times 10^{48}$ & 0.0022 \\
030329 & 0.1685 & 100 & 3    & $1.8\times 10^{49}$ & 0.32   \\
031203 & 0.105  & 100 & 0.35 & $3.1\times 10^{48}$ & 0.048  \\
060218 & 0.0335 & 7   & 0.16 & $2.1\times 10^{46}$ & 0.78   \\
\hline 
\end{tabular}
\end{center}
\end{table}

\end{document}